\documentclass[aps,prd,preprint,superscriptaddress,showpacs,floatfix,preprintnumbers,nofootinbib,longbibliography,a4paper]{revtex4-1}
\usepackage{hyperref}
\usepackage{amsmath}
\usepackage{graphics}
\usepackage{graphicx}
\usepackage{color}
\usepackage{slashed}
\usepackage{subfigure}
\usepackage{amsfonts}
\usepackage{amssymb}
\usepackage{textcomp}
\usepackage[normalem]{ulem}

\hypersetup{
pdfstartview=FitV,
colorlinks=true,
linkcolor=blue, 
citecolor=red, 
filecolor=black, 
urlcolor=blue}

\newcommand*{\be}{\begin{equation}}
\newcommand*{\ee}{\end{equation}}
\newcommand*{\bea}{\begin{eqnarray}}
\newcommand*{\eea}{\end{eqnarray}}

\newcommand{\comment}[1]{}

\newcommand{\cref}[1]{Chapter~\ref{c.#1}}

\def\beq{\begin{equation}}
\def\eeq{\end{equation}}
\def\bea{\begin{eqnarray}}
\def\eea{\end{eqnarray}}
\def\ba{\begin{array}}
\def\ea{\end{array}}
\def\bi{\begin{itemize}}
\def\ei{\end{itemize}}
\def\be{\begin{enumerate}}
\def\ee{\end{enumerate}}
\def\bc{\begin{center}}
\def\ec{\end{center}}
\def\bt{\begin{table}}
\def\et{\end{table}}
\def\btb{\begin{tabular}}
\def\etb{\end{tabular}}

\newcommand{\mgmt}{M_D}

\def\schi{\tilde{\chi}}

\newcommand{\amu}{\delta a_{\mu}}

\begin{document}
\title{Indirect Searches of the Degenerate MSSM}
\author{Debtosh Chowdhury}
\affiliation{Istituto Nazionale di Fisica Nucleare, Sezione di Roma, Piazzale Aldo Moro 2, I-00185,
Rome, Italy.}
\email{debtosh.chowdhury@roma1.infn.it}
\author{Ketan M.  Patel}
\affiliation{Indian Institute of Science Education and Research Mohali, Knowledge City, Sector  81, S A S Nagar, Manauli 140306, India.}
\affiliation{Istituto Nazionale di Fisica Nucleare, Sezione di Padova, I-35131 Padova,
Italy.}
\email{ketan@iisermohali.ac.in}
\author{Xerxes Tata}
\affiliation{Department of Physics and Astronomy,  University of Hawaii, Honolulu, HI 96822, USA.} 
\email{tata@phys.hawaii.edu}
\author{Sudhir K. Vempati}
\affiliation{Centre for High Energy Physics, Indian Institute of Science, Bangalore 560 012,
India.\vspace*{0.5cm}}
\email{vempati@chep.iisc.ernet.in}

\begin{abstract}
A degenerate sfermionic particle spectrum can escape constraints from flavor physics,  and at the same time evade the limits from the direct searches if the degeneracy extends to the gaugino-higgsino sector. Inspired by this, we consider a scenario where all the soft terms have an approximately common mass scale at $M_{\text{SUSY}}$, with splittings $\lesssim \mathcal{O}(10\%)$. As a result, the third generation sfermions have large to maximal (left-right) mixing, the same being the case with charginos and some sectors of the neutralino mass matrix. We study this scenario in the light of discovery of the Higgs boson with mass $\sim$ 125 GeV. We consider constraints from $B$-physics, the anomalous magnetic moment of the muon and the dark matter relic density. We find that a supersymmetric spectrum as light as 600 GeV could be consistent with all current data and also account for the observed anomalous magnetic moment of the muon within $2\sigma$. The neutralino relic density is generally too small to saturate the measured cold dark matter relic density. Direct detection limits from XENON100 and LUX put severe constraints on this scenario which will be conclusively probed by XENONnT experiment.
\end{abstract}
\vspace*{1cm}
\date{\today}
\preprint{UH-511-1273-16}
\maketitle

\section{Introduction}
The first run of the Large Hadron Collider (LHC) at center-of-mass energy $\sqrt{s} = 7$ TeV and 8 TeV has been historic
due to its discovery of a scalar particle of mass close to 126 GeV \cite{Aad:2012tfa,Chatrchyan:2012ufa}. The discovered particle has its properties very
close to the Higgs boson of the Standard Model (SM)
\cite{Khachatryan:2014jba,ATLAS-CONF-2015-007,Aad:2015zhl}. If the nature is supersymmetric, it is
quite likely that the observed  particle would correspond to the lightest CP-even Higgs boson of
the Minimal Supersymmetric Standard Model (MSSM), leading to severe constraints on the MSSM parameter space
\cite{Hall:2011aa,Heinemeyer:2011aa,AlbornozVasquez:2011aa,Arbey:2011aa,Arbey:2011ab,Draper:2011aa,
Carena:2011aa,Christensen:2012ei,Baer:2011ab,Kadastik:2011aa,Buchmueller:2011ab,Aparicio:2012iw,Ellis:2012aa,Baer:2012uya,Arbey:2012dq,Cao:2012fz}. In particular, for stops lighter than 2 TeV,
the stop mixing parameter $X_t$ is required to be as large as $ \sqrt{6} M_{\text{SUSY}}$, with $M_{\text{SUSY}} = \sqrt{m_{\tilde{t}_1} m_{\tilde{t}_2}}$ is the geometric mean of stop masses, leading to a large  stop mixing.
The other alternative is to push the top squarks into the multi-TeV regime, far beyond the reach of LHC.

In addition to the Higgs boson mass measurement, there are two other important sets of constraints on MSSM and
supersymmetry breaking models. One is from  the direct searches for supersymmetric particles at
the LHC. The LHC data have yielded no evidence for supersymmetric particles resulting in a variety of lower bounds on superpartner masses.  These are summarized in various publications by the ATLAS \cite{Aad:2014nra,Aad:2015baa,ATLASCollaboration:2016wlb,Aaboud:2016nwl,Aaboud:2016uth,Aaboud:2016lwz,Aad:2016eki,Aad:2016qqk,Aaboud:2016zdn,Aad:2016tuk,ATLAS-CONF-2016-037,ATLAS-CONF-2016-052,ATLAS-CONF-2016-054,ATLAS-CONF-2016-078} and the CMS \cite{Khachatryan:2016nvf,Khachatryan:2016oia,Khachatryan:2016mbu,Khachatryan:2016pxa,CMS:2016kuv,CMS:2016swi,CMS-PAS-SUS-16-014,CMS-PAS-SUS-16-016,CMS-PAS-SUS-16-022,CMS-PAS-SUS-16-030,CMS-PAS-SUS-16-020,CMS-PAS-SUS-16-019,CMS-PAS-SUS-16-021} collaborations. The LHC constraints are strongest for colored superpartners. Many of these limits are obtained in simplified models where assumptions on mass ordering of SUSY particles are made only on those particles relevant for the particular process.  For a very light neutralino, the limits on gluinos extend to as high as 1.9 TeV. The first generation squarks are also ruled out up to $0.9$-$1.3$ TeV.  The bounds on the second generation squarks could be much weaker if the universality between the first two generations is given up \cite{Mahbubani:2012qq} though one would then have to worry about unwanted flavor effects. Third generation squarks are ruled out up to $900$ GeV for massless Lightest Supersymmetric Particles (LSPs) \cite{Aaboud:2016nwl}. 
Weakly charged particles like neutralinos, charginos and sleptons do not face such strong
constraints from the LHC. For example, the results from chargino pair production and its subsequent
decays exclude the lightest chargino mass up to 100 GeV to 415 GeV for a massless neutralino
\cite{Aad:2014vma}. In scenarios of sleptons decaying into leptons and neutralinos, slepton masses
between 90 GeV and 325 GeV are also excluded if a neutralino is massless \cite{Aad:2014vma}.
However, all these constraints become much weaker when the mass spectrum of neutralinos and charginos is nearly degenerate \cite{Martin:2014qra,Baer:2014kya}. 

The second set of constraints comes from flavor experiments. The $B$-factories and the LHCb experiments have not seen any significant deviations in most rare decay modes of the $B$-mesons from the SM expectations. The $\textrm{BR}(B_s^0 \to \mu^{+} \mu^{-})$ measurement by the LHCb \cite{CMS:2014xfa} sits very close to its SM prediction \cite{Bobeth:2013uxa}. Likewise, the measured values of $\mathrm{BR}(B^{+}\to \tau^{+} \nu_\tau)$ \cite{Abdesselam:2014hkd} and $\mathrm{BR}(B\to X_s \gamma)$ \cite{Lees:2012ufa} are also in agreement with the SM predictions. For the first two generations, the flavor constraints are even more stringent pushing the off-diagonal entries to be smaller than $\mathcal{O}(0.1 \%)$ of the diagonal ones in the sfermion mass matrices.

Assuming supersymmetric particles of SM fermions to be nearly degenerate automatically evades all the flavor constraints. The degeneracy implies a large approximate flavor symmetry which protects flavor violations.  In fact, the scale at which the particles are degenerate can be anything as long as there are no heavy thresholds with large flavor violating couplings between the scale of degeneracy and weak scales. In case of the $R$-parity conserving supersymmetric models, the limits from direct searches at the LHC can also be evaded by assuming a compressed supersymmetric spectrum at the weak scale, see for example \cite{LeCompte:2011cn,LeCompte:2011fh,Dreiner:2012sh,Dreiner:2012gx} for discussions. The deciding factor in this case is the mass difference between the gluino/squark and the produced daughters, and the mass gap(s) between these daughters of the LSP if cascade decays are operative. As long as this difference lies within $100$-$200$ GeV, most direct limits on squarks and gluinos for direct decays to the LSP are inapplicable. For example, in deciding the direct search limits from LHC on the stop, the degeneracy in the stop and neutralino mass plays an important role. Although the lower limits on $m_{\tilde{t}_1}$ extend out to 900 GeV \cite{ATLAS-CONF-2016-077,ATLAS-CONF-2016-050} when the LSP is massless and the top squark decays directly to the LSP, it drops to 400 GeV (300 GeV) if $m_{\tilde{t}_1} - m_{\tilde{\chi}^0_1} < 200~ (100)$ GeV.


In the present work, we extend the hypothesis of degeneracy to all the soft terms and also to the $\mu$ parameter, and consider a Degenerate MSSM (DMSSM). While collider considerations need only partial degeneracy in the full supersymmetric spectra, we extend it to all the sectors thus enabling us to study the constraints from indirect tests, flavors observables and dark matter. In addition, we study the possibility that the DMSSM can provide a solution of the current discrepancy between the SM prediction and the experimental value of muon anomalous magnetic moment. This requires the degenerate scale not to be very high and one obtains an upper bound on sparticle masses\footnote{A similar approach is adopted recently in \cite{Badziak:2014kea} but without assuming degenerate supersymmetry spectra.}. The measured Higgs mass together with the constraints from $B \to X_s \gamma$, on the other hand prefer relatively high degenerate scale. However, we show there still exists a small range for the degenerate scale in $600$-$1000$ GeV for which the total compatibility between the various constraints can be achieved and muon $g-2$ can be brought in agreement with its measured value at the $2\sigma$ level. Such a set-up prefers higher values of $\tan\beta$ and a large trilinear coupling. Further, we find that the LSP cannot make up all of the dark matter of the universe and the present limits on spin-independent neutralino-nucleon cross-section puts severe restrictions on the available parameter space of degenerate spectra.

The paper is organized as follows. In the next section, we briefly review  models in which the weak scale degeneracy arises in the sparticle spectrum. An analytical study of the phenomenological consequences of DMSSM spectrum on muon $g-2$, Higgs mass and some of the flavor observables is presented in section \ref{analytical}. This is then followed by a full numerical analysis in section \ref{numerical}. We then discuss the implications of degenerate SUSY spectra Ïfor dark matter in section \ref{dm}. Finally, we summarize in section \ref{summary}.

\section{SUSY Models with weak scale degeneracy}
\label{models}
The near degeneracy in sparticle spectrum is seen as one of the explanations which allow for low-energy supersymmetry, given the absence of its signal at the LHC so far. While this option has been widely studied phenomenologically (see for example, \cite{LeCompte:2011cn,LeCompte:2011fh,Dreiner:2012sh,Dreiner:2012gx,Barducci:2015ffa,An:2015uwa}), its theoretical justifications based on the explicit models of supersymmetry breaking are very limited. A supersymmetry breaking model for nearly degenerate sparticles at the weak scale would have the following characteristics:
\begin{itemize}
\item  One would naively expect it be a low scale mediation model, since we are demanding
degeneracy of the soft terms at the weak scale. Any degeneracy from a high scale mediation model
would be lost by the renormalization group evolution which introduces large non-degeneracy at least
between colored and uncolored superpartners while running from the high scale to the weak scale.
\item In some cases, partial weak scale degeneracy can arise from special high scale scenarios
such as in the models based on mixed moduli-anomaly mediation
\cite{Choi:2005ge,Choi:2005uz,Falkowski:2005ck,Baer:2006id,Baer:2006tb,Kawagoe:2006sm,Baer:2007eh}.
\item The model is also required to be flavor universal except for Yukawa couplings effects. While this may be automatically
arranged in the models of universal masses, it can also be implemented by explicit imposition of
flavor symmetry at the weak scale in the soft breaking sector.
\item The $\mu$ parameter is required to be very close to the gaugino soft masses in order to keep all the charginos and neutralinos approximately degenerate in their masses. 
\end{itemize}

An interesting class of models in which some of these features can be realized arise from the supersymmetry breaking by compactification, with twisted boundary conditions, of an extra spatial dimension, also known as the Scherk-Schwarz mechanism \cite{Scherk:1979zr}. In the simplest version, an extra dimension is compactified on an orbifold, $S^1/Z_2$, and 5D $N=1$ supersymmetry is completely broken on one of the two branes by the combined action of $Z_2$ and non-trivial twists \cite{Pomarol:1998sd,Barbieri:2001yz}. The gauge fields, matter fields and Higgs live in the bulk and the $\mu$ term is forbidden by a global $SU(2)_H$ and orbifold symmetry. The $Z_2$ symmetry of an orbifold breaks 5D $N=1$ supersymmetry down to the 4D $N=1$ supersymmetry on the branes. $N=1$ supersymmetry and $SU(2)_H$ are then broken by the non-trivial twists which are parametrized by $\alpha$ and $\gamma$, both less than unity, for matter and Higgs fields respectively. As a result of this, all the MSSM soft parameters can be obtained as functions of only three free real parameters, namely $\alpha$, $\gamma$ and the compactification scale $\sim 1/R$. At the tree level, they are given as
\cite{Barbieri:2001yz}:
\beq \label{ss1}
M_1 = M_2 = M_3
=\frac{\alpha}{R},~~m^2_{H_u}=m^2_{H_d}=m^2_{\tilde{Q}}=m^2_{\tilde{U}}=m^2_{\tilde{D}}=
m^2_{\tilde{L}}=m^2_{\tilde{E}}=\left( \frac{\alpha}{R} \right)^2 \eeq
\beq \label{ss2}
A=-3 \frac{\alpha}{R},~~\mu = \frac{\gamma}{R},~~\mu B = - 2\frac{\alpha \gamma}{R^2}\, , \eeq
where $M_{1,2,3}$ are gaugino mass parameters and $m_\phi$ are the soft masses of various scalars in the MSSM. Here $\alpha$ and $\gamma$ are real parameters and the soft masses are flavor universal as the geometry does not distinguish between the flavors. As a result, the above spectrum naturally solves the flavor and CP problems. The large trilinear coupling predicted by the model favors the observed large Higgs mass. The $\mu$ parameter coincides with the supersymmetry breaking scale if $\alpha \approx \gamma$. The SUSY breaking scale is characterized by $\alpha/R$ which can be different from the compactification scale $1/R$. The renormalization group evolution effects remain small as long as $1/R$ is not well beyond the TeV scale. Further, the radiative corrections to the above masses at and above the compactification scale are under control and are naturally small because of the symmetries of higher spacetime. Hence the above spectrum possesses all the features listed earlier in this section.

There exists other variants of this framework which also lead to approximate degenerate spectrum for sparticles. For example in \cite{Murayama:2012jh}, the Higgs multiplets were localized
on the branes resulting $m^2_{H_u}=m^2_{H_d}=0$, $A=-2 \alpha /R$, $B=0$ at the tree level and
leaving $\mu$ as a free parameter. The $m^2_{H_u}$ and $m^2_{H_d}$ are then generated radiatively which triggers electroweak symmetry breaking. In the more predictive models of similar kind, one can also fix $\alpha =1/2$ by considering an additional $Z_2$ symmetry of an orbifold, {\it i.e.} the
extra-dimension compactified on $S^1/(Z_2 \times Z_2')$ \cite{Barbieri:2000vh,Barbieri:2001dm,Barbieri:2002sw}. 

Low energy degeneracies in supersymmetric spectrum are also realized in mixed moduli-anomaly  mediated models \cite{Choi:2005ge,Choi:2005uz,Falkowski:2005ck,Baer:2006id,Kawagoe:2006sm}. 
These models are realized from string compactifications of Type II B on complex spaces like Calabi-Yau  as in the Kachru, Kallosh, Linde and Trivedi (KKLT) \cite{Kachru:2003aw} setup. Specific regions of parameter spaces in these models, where the splittings at the high scale are compensated by renormalization group evolution, result into a low scale degeneracy as has been emphasized in  \cite{Kawagoe:2006sm}, see also \cite{Bae:2007pa}. We remark, in passing, that the sub-TeV sparticle spectrum that we suggest may bring the value of the muon anomalous magnetic moment in accord with bounds from the LHC and low energy data will not be in strong conflict with SUSY providing the resolution of the naturalness problem.

In the present work, our approach is completely phenomenological. We are driven only by the data to consider the unconventional possibility that all superpartners are approximately degenerate. Although there are top-down mechanisms that lead to a high degree of degeneracy for some (or even the most) superpartners, we recognize that assuming all sparticle to have their masses within narrow range will require explanation. That said, given that we really have no compelling mechanism for how superpartners acquire masses, or how the $\mu$ parameters is generated, we felt that an examination of the observable consequences of any viable framework, no matter how unorthodox, is warranted. With this in mind, we consider the possibility that all soft SUSY-breaking mass terms as well as $|\mu|$ to assume nearly the same value, within $\pm 10 \%$, at the weak scale. We make no representation as to how such a degeneracy might occur, and leave the overall scale of degeneracy as a free parameter.

\section{Degenerate MSSM: An analytic study}
\label{analytical}
In the following, we examine the DMSSM model as a solution of the
muon anomalous magnetic moment discrepancy in the light of measured Higgs mass and updated limits on the most relevant $B$-physics observables using simplified analytical formulae. We begin by summarizing some important aspects of degenerate soft mass parameters on the physical mass spectrum of the MSSM. In our definition of DMSSM, we set the following soft masses at the weak scale to be degenerate with a common scale, namely
\beq \label{deg}
M_{1}\approx M_{2}\approx M_{3} \equiv M_D,~~~
m^2_{\tilde{Q}}\approx m^2_{\tilde{U}}\approx m^2_{\tilde{D}}
\approx m^2_{\tilde{L}} \approx m^2_{\tilde{E}} \equiv M_D^2~. \eeq
Since our approach is more phenomenological and we do not rely on the specific models of SUSY
breaking, we consider $\mu$ and pseudoscalar Higgs mass $m_A$ as free parameters instead of fixing
them in terms of $m_{H_u}$ and $m_{H_d}$. We define 
\beq \label{deg-ewsb}
|\mu|^2 = k_{\mu}~M_D^2,~~~{\rm and}~~~ m_A^2 = k_A~M_D^2, \eeq
where $k_\mu$ and $k_A$ are real and positive parameters of ${\cal O}(1)$. 
Some specific choice of $\alpha$ and $\gamma$ parameter together with
appropriate radiative corrections can lead to correct electroweak symmetry breaking and $|\mu|$ and
$m_A$ as written in Eq.~\eqref{deg-ewsb} \cite{Barbieri:2001yz}. Eq.~\eqref{deg-ewsb} then determines the tree level masses of physical scalars in the Higgs sectors which are given as $m^2_{h} \approx m_Z^2 \ll M_D^2$ and $m_{H}^2 \approx m_{H^\pm}^2 \approx m_A^2 \approx k_A~M_D^2$. The well-known MSSM radiative corrections discussed below would then raise the $m_{h}$ to its observed value.

Next, let us consider the chargino and neutralino mass spectrum in the limit defined in Eqs.~(\ref{deg},\ref{deg-ewsb}). The neutralino mass matrix leads to two physical states with mass $\sim M_D$ with negligible mixing among them ({\it i.e.} pure bino and wino like) and other two states (mostly Higgsino-like) with mass $\sim |\mu|$ which are maximally mixed in the limit $|\mu|,~M_D \gg m_Z$. All four states turn out to be nearly degenerate only if $k_\mu \approx 1$ in Eq.~\eqref{deg-ewsb}. In the same limit, one also gets approximately degenerate charginos \cite{Martin:1997ns}.  Since we are interested in fully compressed sparticle spectrum, we consider $|\mu|$ to be degenerate with $M_D$, or equivalently $k_\mu \approx 1$, in our study of DMSSM. Note that the degeneracy between all the electroweak gauginos get removed when $M_D$ is close to $m_Z$ with splittings that depend on $\tan\beta$. The gluinos also remain degenerate with other gauginos as enforced by condition Eq.~\eqref{deg}. We emphasize that we only mean an approximate degeneracy in the masses of sparticles. In the numerical analysis of the next section, we allow small splittings, and assume that the lightest neutralino is indeed the lightest sparticle.

The masses of the first two generations of squarks and sleptons are almost degenerate in this limit and are of the order of $\sim M_D$. The masses of third generation sfermions receive significant correction from the trilinear terms and in the degenerate soft mass limit their mass matrices can be written as 
\begin{equation} \label{3gen-squark}
m_{\tilde{f}}^2 = \left ( \begin{array}{cc}  M_D^2+m_f^2+\Delta_{\tilde{f}_L} & m_f X_f \\
m_f X_f & M_D^2+m_f^2+\Delta_{\tilde{f}_R}  \end{array} \right)~,
\end{equation}
where $f=t,b,\tau$; $X_t=A_t- \mu \cot\beta$ and $X_{b,\tau}=A_{b,\tau}- \mu \tan\beta$. The
$\Delta_{\tilde{f}_{L,R}}$ represents a contribution to the squarks and slepton masses from the
electroweak symmetry breaking which is negligible when $M_D>m_Z$ \cite{Martin:1997ns}.
Eq.~\eqref{3gen-squark} automatically leads to large mixing between the stops which
is favored by the large Higgs boson mass. The splittings between the sfermions of a given SM charge
is given by $m_{\tilde{f_2}}^2 - m_{\tilde{f_1}}^2 \approx |2  m_f X_f| $ which leads to the largest
deviation from the degeneracy in the stop sector. The scale of supersymmetry breaking, defined as
geometric mean of stop masses, is given by
\beq \label{msusy-def}
M_{\rm SUSY} \equiv \sqrt{m_{\tilde{t}_1} m_{\tilde{t}_2}} = M_D \left( 1- \frac{m_t^2
X_t^2}{M_D^4} + 2\frac{m_t^2}{M_D^2} + \frac{m_t^4}{M_D^4} \right)^{1/4}~.\eeq
In the numerical analysis, we have also considered the full one loop radiative corrections on all sfermion mass matrices. These corrections can play an important role especially in the limit of large $m_A$ as will be elaborated further. We now discuss below the impact of such a degenerate mass spectrum on the various observables.

\subsection{The Higgs boson mass}
Next we turn to constraints on the degenerate scale $M_D$ from the measurement of Higgs mass. At the tree level, as usual the Higgs spectrum is fixed by $m_A$ ($\sim k_{A} M_D$ by Eq.~\eqref{deg-ewsb}) and $\tan\beta$. Radiative corrections are extremely important. In the MSSM, the one loop corrected lightest CP-even Higgs mass can be expressed as
\cite{Altmannshofer:2012ks}: 
\begin{eqnarray} \label{higgsmass}
m_h^2 &=& m_Z^2 \cos^2 2 \beta + \delta m_h^2~,  \nonumber \\
\delta m_h^2 &=&  {3 \over 4 \pi^2} { m_t^4 \over v^2} \left( \log \left( M_{\rm SUSY}^2 \over
{m_t^2} \right)  + {X_t^2 \over M_{\rm SUSY}^2 } - {X_t^4 \over 12 M_{\rm SUSY}^4} \right)
\nonumber \\
&-& {3 \over 48 \pi^2} {m_b^4 \over v^2}  {\tan\beta^4 \over ( 1+ \epsilon_b \tan\beta)^4 } {\mu^4
\over m_{\tilde{b}}^4} \nonumber \\
&-& {1 \over 48 \pi^2} {m_\tau^4 \over v^2} { \tan\beta^4 \over ( 1+ \epsilon_l \tan\beta)^4 }
{\mu^4 \over m_{\tilde{\tau}}^4}~,
\end{eqnarray}
where $M_{\rm SUSY}$ is given in Eq.~\eqref{msusy-def} while  $m_{\tilde{b}}$ and $m_{\tilde{\tau}}$ are average sbottom and stau mass and can be identified with $M_D$ in the degenerate limit. The
$\epsilon_i$ factors arise from the corrections to the Higgs-fermion-fermion couplings and their complete expressions are given in \cite{Altmannshofer:2012ks}. We use the above formula to estimate
the Higgs mass in the limit defined in Eq.~\eqref{deg}. In this
limit the $\epsilon_i$ are given by
\begin{eqnarray}
\epsilon_b &=& \epsilon_b^{\tilde{g}} + \epsilon_b^{\tilde{W}} + \epsilon_b^{\tilde{H}}~, \nonumber
\\
\epsilon_b^{\tilde{g}} &=& {\alpha_s \over 3 \pi} ~{\mu \over M_D}, \nonumber \\
\epsilon_b^{\tilde{W}} &=& - {\alpha_2 \over 4 \pi} ~{3 \over 2} ~\mu~ M_D ~\tilde{g}(\mu^2,
M_D^2) \simeq -\frac{3 \alpha_2}{16 \pi},  \nonumber \\
\epsilon_b^{\tilde{H}} &=&  - {\alpha_2 \over 4 \pi} ~{m_t^2 \over 2 M_W^2} ~\mu~ A_t~ \tilde{g}(\mu^2, M_D^2) \simeq -\frac{\alpha_2}{16 \pi} \frac{m_t^2}{M_W^2} \frac{A_t}{M_D}, \nonumber \\
\epsilon_l &=& - {\alpha_2 \over 4 \pi} ~{3 \over 2} ~\mu~ M_D ~\tilde{g}(\mu^2, M_D^2) \simeq -\frac{3 \alpha_2}{16 \pi}  ,
\end{eqnarray}
and the function $\tilde{g}(x,y)$ is given by
\begin{equation}
\tilde{g}(x,y) = {-x+y+x\log(x)-x\log(y)\over(x-y)^2} \Rightarrow \lim_{y\rightarrow x}
\tilde{g}(x,y) = {1 \over 2x}.
\end{equation}
As already mentioned, we also take $\mu\approx M_D$ limit while estimating the Higgs mass using Eq.~\eqref{higgsmass}. The combined experimental measurements of the Higgs mass by CMS and ATLAS \cite{Aad:2015zhl} allow a window of $124.4-125.8$ GeV at $3\sigma$. In addition, there is theoretical uncertainty in Higgs mass calculation, owing to uncertainty in top quark mass determination, scheme dependence, residual three loop effects etc. (see \cite{Vega:2015fna} for a discussion). Considering this, we also allow additional $\pm 2$ GeV uncertainty in the Higgs mass to account for these theoretical uncertainties. Hence, the conservative Higgs mass range considered by us is $122.4-127.8$ GeV.

\subsection{The anomalous magnetic moment of muon} 
\begin{figure}[t]
\centering
\subfigure{\includegraphics[width=0.47\textwidth]{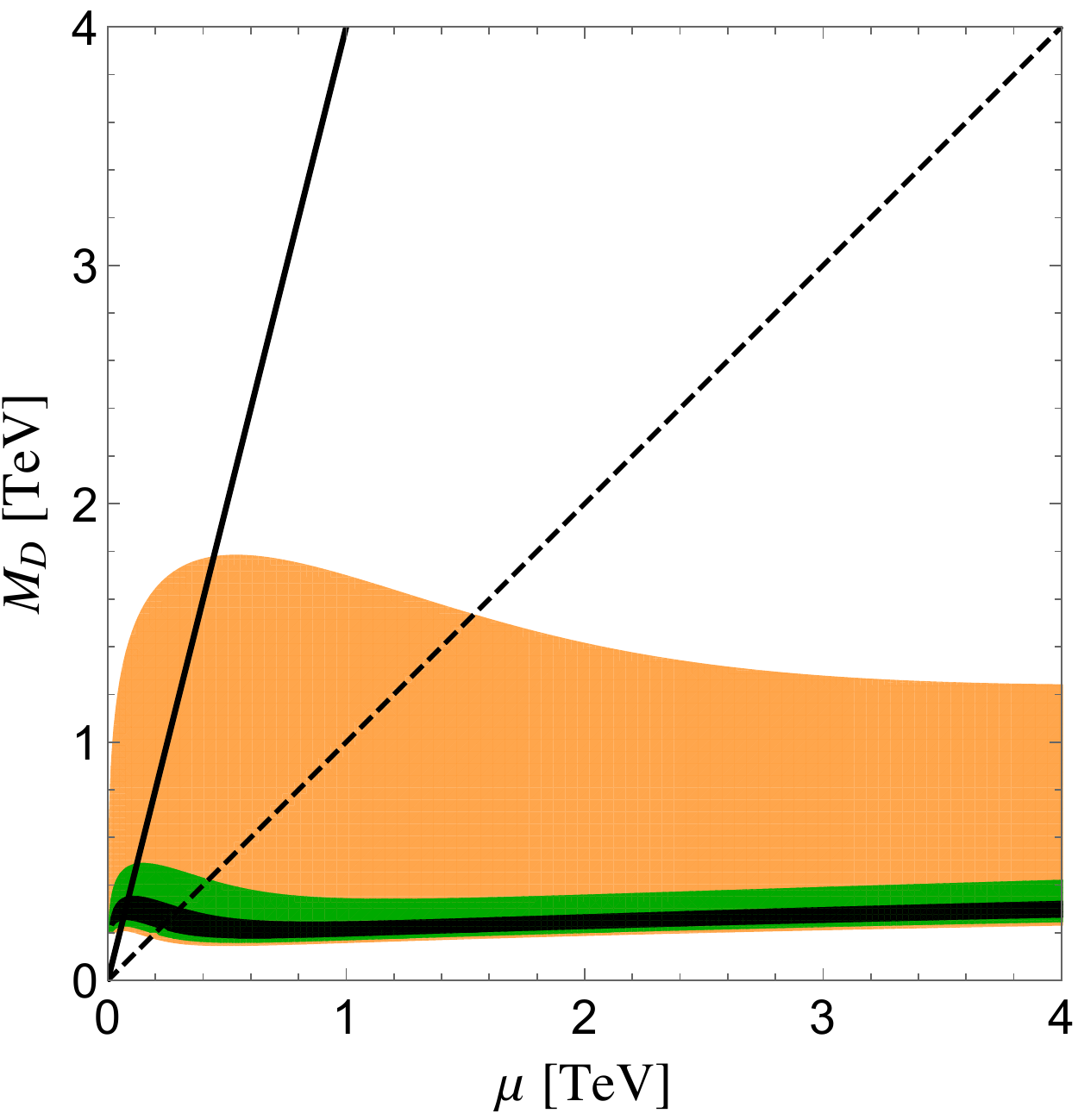}}\quad
\subfigure{\includegraphics[width=0.47\textwidth]{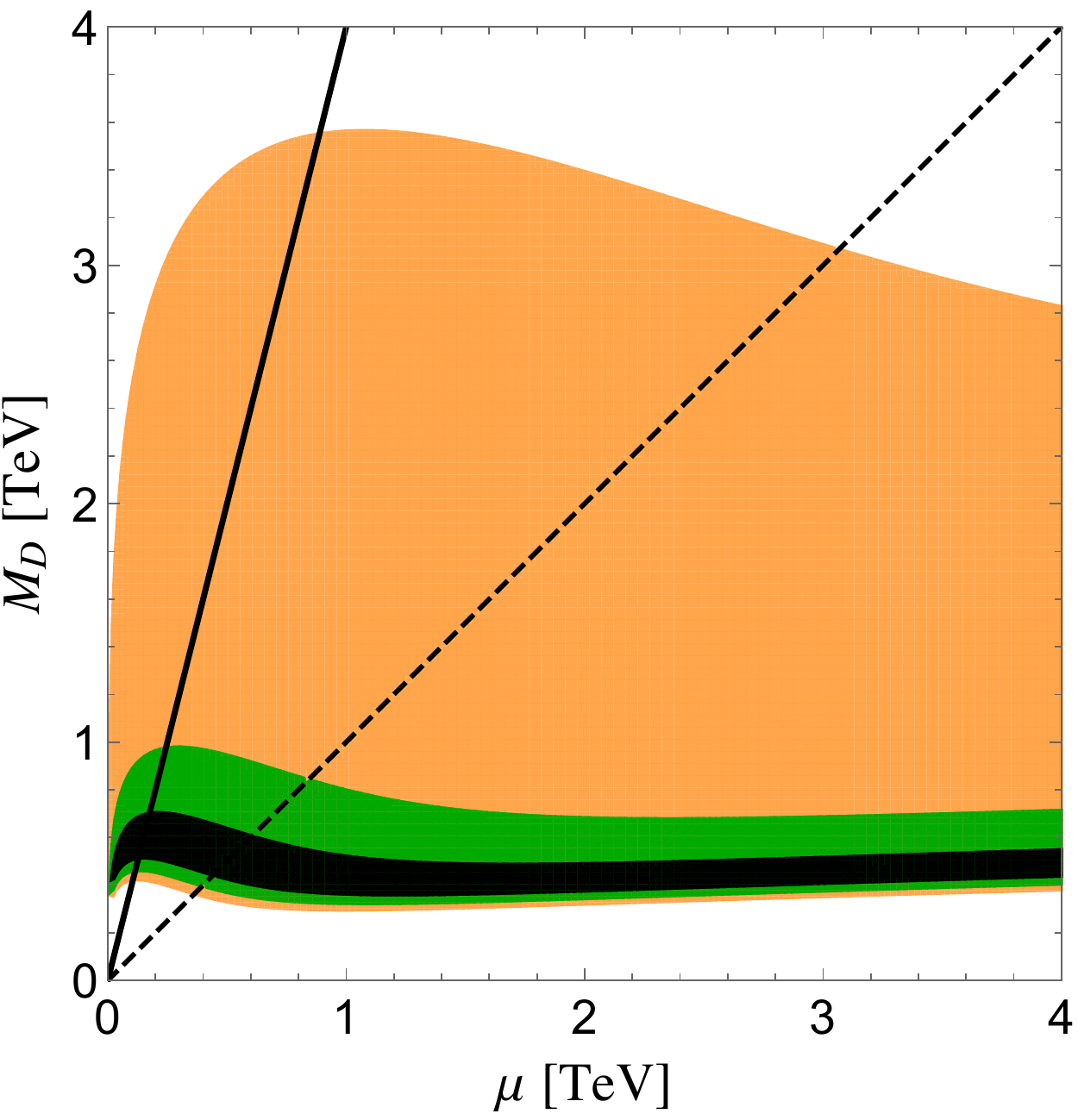}}
\caption{The parameter space allowed in $\mu$-$M_D$ plane by the $1\sigma$ (black), $2\sigma$
(green) and $3\sigma$ (orange) ranges of the $\delta a_\mu$ for $\tan\beta=10$ (left
panel) and $\tan\beta=40$ (right panel). The solid and dashed lines correspond to
$\mu=\frac{1}{4}M_D$ and $\mu=M_D$, respectively.}
\label{fig1}
\end{figure}

We require the degenerate scale to be low enough to resolve the current $a_\mu=(g-2)_\mu/2$
discrepancy. In the MSSM, the leading contributions to $\delta a_\mu$ at one loop comes from the
chargino and bino exchanges and are given by
\cite{Giudice:2012pf}
\begin{eqnarray}
\delta a_\mu  &=&  \frac{\alpha ~m_\mu^2~ \mu~ M_2 \text{tan} \beta}{4 \pi \sin^2 \theta_W m_L^2} 
        \left({ f_\chi[M_2^2/m_L^2] - f_\chi[ \mu^2/m_L^2] \over (M_2^2 - \mu^2 )}\right) \nonumber
\\
        &+& 
        {\alpha ~m_\mu^2~ \mu M_1 \tan\beta \over 4 \pi \cos^2\theta_W (m_R^2 - m_L^2)} \left( 
        {f_N[M_1^2/m_R^2] \over m_R^2} -  {f_N[M_1^2/m_L^2] \over m_L^2} \right), 
\end{eqnarray}      
where, 
\begin{eqnarray}
f_\chi[x] &=& {x^2 - 4 x + 3 + 2 \ln (x) \over (1-x)^3} \Rightarrow \lim_{x\rightarrow 1}f_\chi[x] =
-\frac{2}{3}, \nonumber  \\
f_N[x] &=& { x^2 -1 - 2 x \ln(x) \over (1 -x)^3}\Rightarrow \lim_{x\rightarrow 1}f_N[x] =
-\frac{1}{3}.
\end{eqnarray}
In the degenerate limit defined by Eq.~\eqref{deg}, we have  
\begin{eqnarray}
\delta a_\mu 
      &=&  { \alpha~ m_\mu^2~ \mu~  \tan\beta  \over 4 \pi \sin^2\theta_W M_D^3 }
      \left( { 1 \over (1 -x_r^2)} \left( - {2 \over 3} - f_\chi[x_r^2] \right) + {1 \over 3
\cot^2\theta_W } \right),  \end{eqnarray}
where $x_r \equiv \mu/M_D$. In the limit when $x_r \approx 1$, the first term in the parenthesis becomes $\approx 1/4$ and dominates over the second one. The current discrepancy between the SM calculation and experimental measurement (see \cite{Marciano:2016yhf} and references therein) of $a_\mu$ is
\beq \label{g-2-exp}
\delta a_\mu = a_\mu^{\text{exp}} - a_\mu^{\text{SM}} = (2.73 \pm 0.80) \times 10^{-9}, \eeq
which is about $3.4\sigma$ deviation from the SM value. An alternative analysis which gives about $4\sigma$ deviation can be found in \cite{Jegerlehner:2015stw}. The region in $\mu$ and $M_D$ allowed to resolve the above discrepancy in $\delta a_\mu$ is shown in Fig.~\ref{fig1} for two specific values of $\tan\beta$.

We see that, for a large range of $\mu$, the degenerate scale $M_D$ in the 200-500 (400-1000) GeV range can account for the $(g-2)_\mu$ discrepancy at $2\sigma$ for small (large)
$\tan\beta$. Note that for a given scale $M_D$, the $\delta a_\mu$ picks up the maximum
value when $\mu \approx \frac{1}{4}M_D$. However, in the $\mu \ll M_D \sim M_{1,2}$ limit,
the lightest neutralino becomes Higgsino-like and the degeneracy between the lightest neutralino and
sfermions gets destroyed. One is forced to take $\mu \approx M_D$ if we want to have an almost degenerate spectrum of charginos and neutralinos in order to be safe from the LHC constraints.

\subsection{$B \to X_s \gamma$ and $B_s \to \mu^+ \mu^-$}
Before we present the results of Higgs mass constraints on $M_D$ and $A_t$, let us also consider the relevant constraints coming from the branching fraction for the $B \to X_s \gamma$ and $B_s \to \mu^+ \mu^-$ that are sensitive to new physics. In the MSSM the contribution to $B \to X_s \gamma$ is given by
\cite{Misiak:2015xwa}
\begin{eqnarray} \label{bsg}
R_{bs\gamma}&\equiv&{\text{BR}(B \to X_s \gamma) \over \text{BR}(B \to X_s \gamma)_{\text{SM}}} =1-2.45
C_7^{NP}-0.59C_8^{NP},
\end{eqnarray}
where $C_{7,8}^{NP}$ are Wilson coefficients which encode the new physics contributions to the magnetic and chromo-magnetic $b \to s \gamma $ operators and their most general expressions in the MSSM case are given in \cite{Altmannshofer:2012ks}\footnote{The coefficients in Eq.~\eqref{bsg} are taken from the updated theoretical prediction given in Eq.~(10) of \cite{Misiak:2015xwa} and then divided by the SM central value to get the ratio given in Eq.~\eqref{bsg}. When writing Eq.~(10) of \cite{Misiak:2015xwa}, it is assumed that the quadratic terms are negligible when $C_7$ and $C_8$ enter in Eq.~\eqref{bsg} with $\mathcal{O}(1)$ coefficients.}. In the case of degenerate soft masses and $\mu \simeq M_D$, they can be written as:
\begin{eqnarray} \label{c7c8}
C_{7,8}^{NP} &=& C_{7,8}^H+C_{7,8}^{\tilde{H}}+C_{7,8}^{\tilde{W}}+C_{7,8}^{\tilde{g}}~,\nonumber \\
C_{7,8}^{H}&=&\left( {1-\epsilon_0 t_\beta \over 1+\epsilon_b t_\beta}+{(\epsilon_b^{\tilde{H}})^2
t_\beta^2 \over (1+\epsilon_bt_\beta)(1+\epsilon_0t_\beta)}\right){m_t^2
\over{2m_{H^+}^2}}h_{7,8}\left({m_t^2 \over m_{H^+}^2} \right)  
\nonumber \\
&+&{\epsilon_b^{\tilde{H}} t_\beta^3 \over (1+\epsilon_bt_\beta)^2(1+\epsilon_0t_\beta)}{m_b^2 \over
2 m_A^2}z_{7,8}, ~ \nonumber \\
C_{7}^{\tilde{H}}&=&-{t_\beta \over 1+\epsilon_b t_\beta}{5 \over 72}{A_t m_t^2 \over
M_D^3},~~C_{8}^{\tilde{H}}={3\over 5}C_{7}^{\tilde{H}},~ \nonumber \\
C_{7}^{\tilde{g}}&=&{g_3^2 \over g_2^2}{\epsilon_b^{\tilde{H}} t_\beta^2 \over
(1+\epsilon_bt_\beta)(1+\epsilon_0t_\beta)}{2 \over 27}{m_W^2 \over
M_D^2},~~C_{8}^{\tilde{g}}={15\over 4}C_{7}^{\tilde{g}},~ \nonumber \\
C_{7}^{\tilde{W}}&=&{\epsilon_b^{\tilde{H}} t_\beta^2 \over
(1+\epsilon_bt_\beta)(1+\epsilon_0t_\beta)}{7\over 24}{m_W^2 \over
M_D^2},~~C_{8}^{\tilde{W}}={3\over 7}C_{7}^{\tilde{W}},~
\end{eqnarray}
where and $m_{H^\pm}^2=m_A^2+m_W^2$, $\epsilon_0=\epsilon_b-\epsilon_b^{\tilde{H}}$, $z_7=-1/18$, $z_8=1/6$ and function $h_{7,8}(x)$ are given in the Appendix in Ref. \cite{Altmannshofer:2012ks}. In the SM, the NNLO prediction for the branching ratio is BR$(B \to X_s \gamma)_{\text{SM}} = (3.36 \pm 0.23) \times 10^{-4}$ \cite{Czakon:2015exa,Misiak:2006zs} while the present world average of experimental measurements reads BR$(B \to X_s \gamma)_{\text{exp}} = (3.49 \pm 0.19) \times 10^{-4}$ \cite{Amhis:2014hma}. This leaves the following room for new physics in the $R_{bs\gamma}$ defined in Eq.~\eqref{bsg}
\beq \label{Rbsg}
R_{bs\gamma}=1.04 \pm 0.09~. \eeq

We also calculate the new physics contribution to BR$(B_s\to \mu^+\mu^-)$ using the estimation given in \cite{Altmannshofer:2012ks}. The MSSM contribution to this leptonic decay can be approximated in
the large $\tan\beta$ limit as
\beq \label{Rbsmm}
R_{B_s\mu\mu} \equiv \frac{\text{BR}(B_s \to \mu^+ \mu^-)}{\text{BR}(B_s \to
\mu^+ \mu^-)_{\text{SM}}} \simeq |{\cal A}|^2+|1-{\cal A}|^2, \eeq
where an updated calculation of BR$(B_s\to \mu^+\mu^-)$ in the SM gives
\cite{Bobeth:2013uxa}
\beq \label{RbsmmSM}
\text{BR}(B_s \to \mu^+ \mu^-)_{\text{SM}} = (3.65 \pm 0.23)\times 10^{-9}~. \eeq
${\cal A}$ contains the new contribution which is mainly due to the exchanges of heavy
neutral Higgs and pseudoscalar Higgs with their flavor changing couplings induced at one loop. 
It is parametrized as
\beq \label{RbsmmA}
{\cal A} = \frac{4\pi}{\alpha_2} \frac{m^2_{B_s}}{4 M_A^2}\frac{\epsilon_{FC}
\tan^3\beta}{(1+\epsilon_b \tan\beta)(1+\epsilon_0 \tan\beta)(1+\epsilon_l \tan\beta)}
\frac{1}{2 C_A}, \eeq
where $\epsilon_i$ are already specified above in the degenerate limit while $C_A$ is SM loop
function and is approximately given as $C_A \simeq 0.469$ \cite{Bobeth:2013uxa}. The new contribution in this leptonic decay strongly depends on the pseudoscalar Higgs mass, $m_A$. For definiteness, we use $m_A = M_D$ in this analysis. Note however that one can even use $m_A > M_D$ (without affecting the other phenomenology, in particular the direct LHC constraints) as it is allowed by Eq.~\eqref{deg-ewsb} and can thus evade the constraints from $B_s\to \mu^+\mu^-$. However, for the analysis presented in this section we take $m_A=M_D$ to estimate the constraints on $A_t$ and $M_D$. The combined analysis from LHCb and CMS imply ${\rm BR}(B_s \rightarrow \mu^+ \mu^-)  \in 2.8^{+0.7}_{-0.6} \times 10^{-9}$ \cite{CMS:2014xfa} resulting into a constraint, $0.37 < R_{B_s\mu\mu} < 1.17$ at $2\sigma$.

\begin{figure}[t]
\centering
\subfigure{\includegraphics[width=0.47\textwidth]{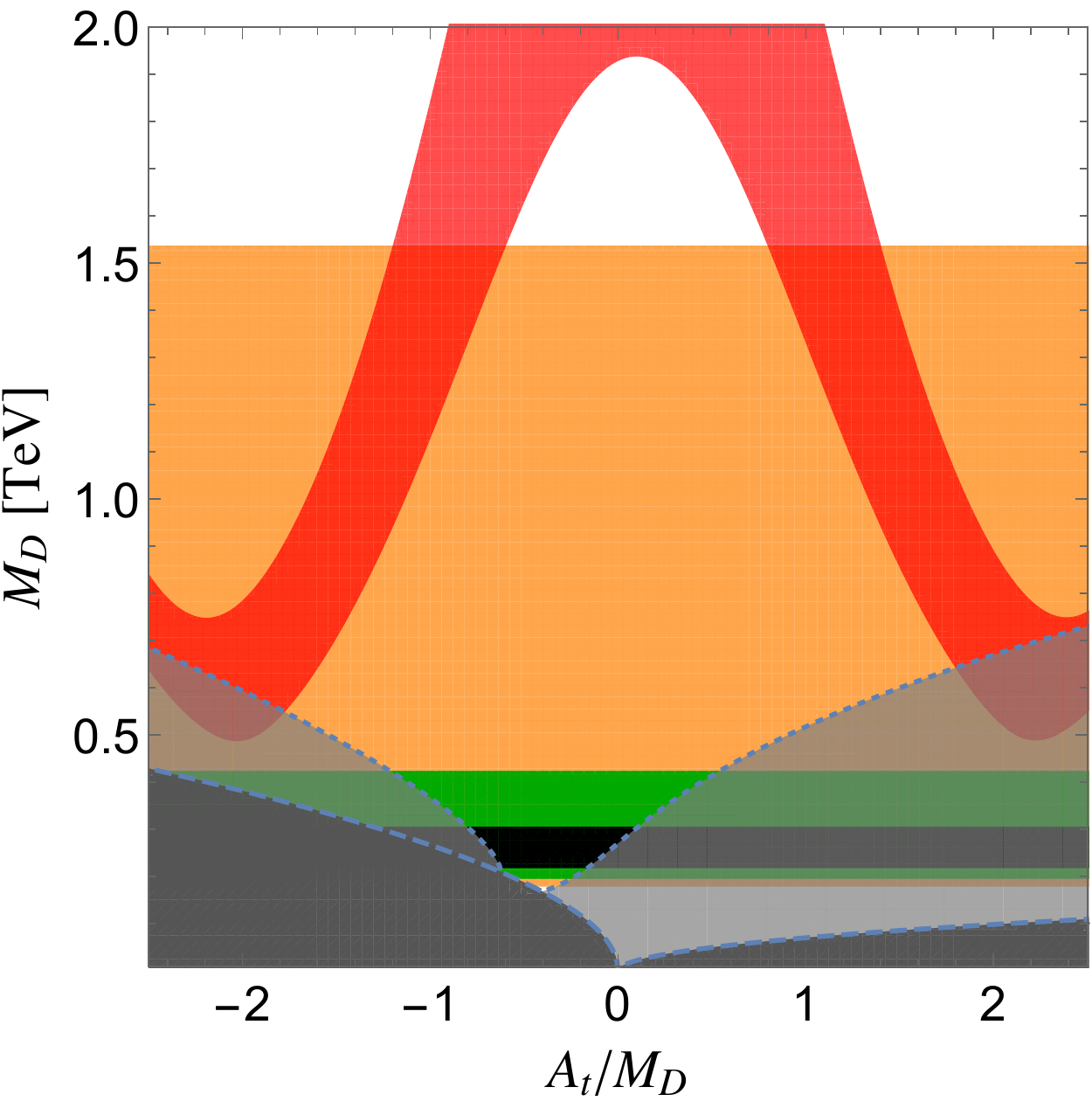}}\quad
\subfigure{\includegraphics[width=0.47\textwidth]{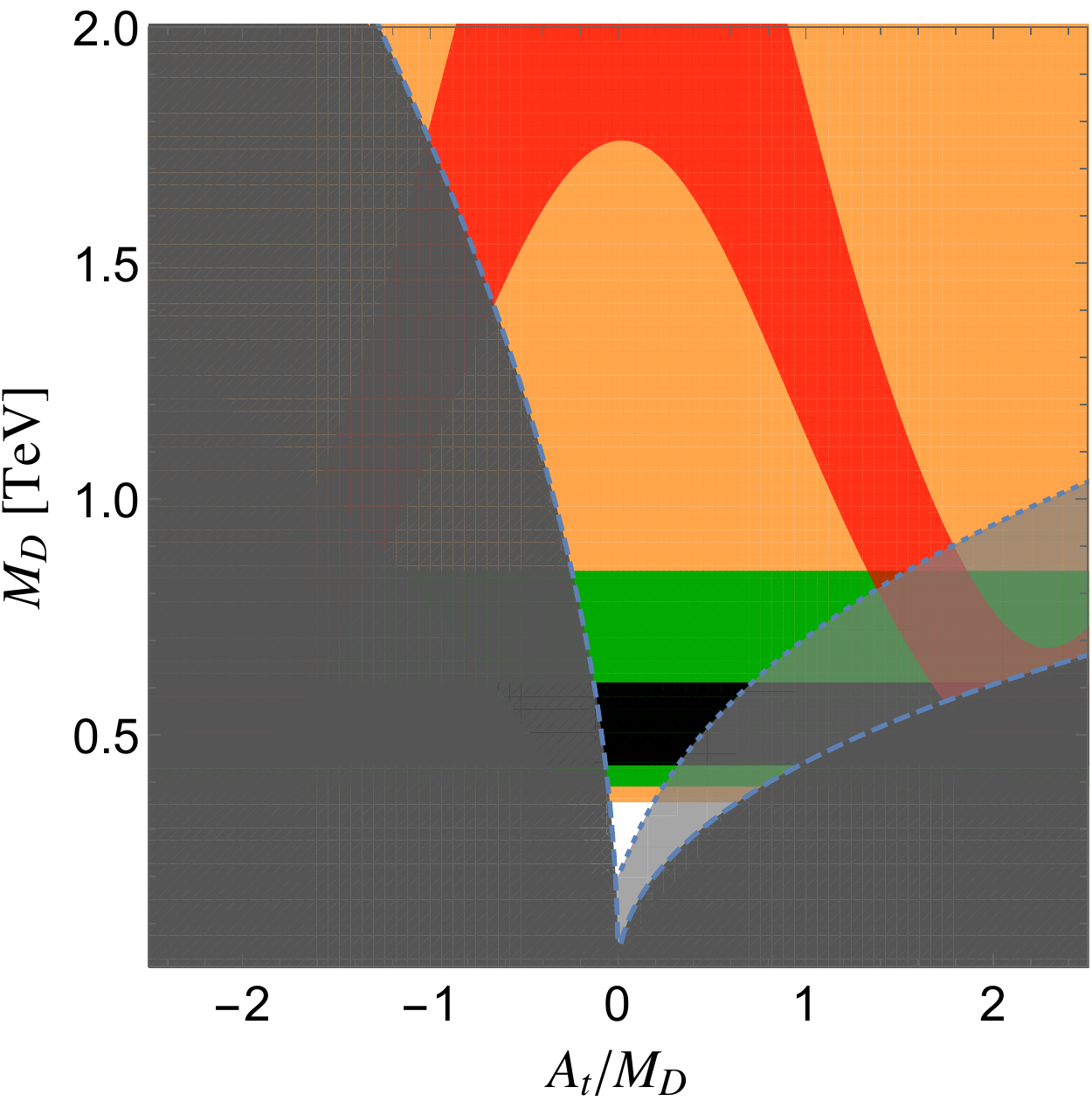}}
\caption{The parameter space allowed in $A_t$-$M_D$ plane by different constraints for
$\tan\beta=10$ (left panel) and $\tan\beta=40$ (right panel). The horizontal black, green and orange
bands show the favored values of $M_D$ by $\delta a_\mu$ at $1\sigma$, $2\sigma$ and $3\sigma$
respectively. The red band corresponds to a valid Higgs mass (122.4-127.8 GeV) region. The lighter
(darker) gray region is excluded by BR$(B \to X_s \gamma)$ (BR$(B_s\to \mu^+\mu^-)$) at $2\sigma$.}
\label{fig2}
\end{figure}

The results for the semi-analytical computation for various observables, just discussed are displayed in Fig.~\ref{fig2}. We see that:
\begin{itemize}
 \item A relatively large $A_t$ is essential to generate the correct Higgs mass for the low values
of $M_D$ which is required to produce sizable $\delta a_\mu$.
 \item On the other hand, cancellation in the flavor violating effects due to degenerate SUSY
spectrum works well only for vanishing $A_t$. The SUSY contribution to $C_{7,8}$ by Higgsinos,
gluinos and charginos vanishes if $A_t\approx 0$ as can be seen from Eq.~\eqref{c7c8}. For nonzero
$A_t$ 
the process mediated by Higgsino-stop loop dominate in this case and leads to large flavor violating effects. This puts severe constraints on the allowed regions.
\item The contribution in $B\to X_s \gamma$ due to the charged Higgs loop, namely $C_{7,8}^H$, is positive. As can be seen from Eq.~\eqref{c7c8}, the Higgsino-stop loop also contributes through $C_{7,8}^{\tilde{H}}$ positively for negative $A_t$ leading to large $B\to X_s \gamma$. This disfavors negative $A_t$. 
\item For small values of $\tan\beta$, relatively low $M_D \in
[200-500]$ GeV is required to bring $\delta a_\mu$ in $2\sigma$ agreement with the observed value.
However such a low $M_D$ is disfavored by both the observed Higgs mass and $B \to X_s
\gamma$ constraints. 
\item Large $\tan\beta$ allows one to increase $M_D$ so that the Higgs mass and $B \to X_s
\gamma$ constraints can simultaneously be satisfied but only in a tiny region around $A_t/M_D \approx 1.5$.  In this case, the $\delta a_\mu$ can be brought
in to the 95\% C.L. agreement with its observed value being  consistent with the Higgs mass and
$2\sigma$ limits on $B \to X_s \gamma$ and $B_s\to \mu^+\mu^-$.
\end{itemize}

This simplified analytical study indicates that there is room for a phenomenological viable solution of $(g-2)_\mu$ discrepancy at the $2\sigma$ level, though such solution requires relatively large degeneracy scale $M_D \ge 800$ GeV which appears to be beyond the present reach of LHC because of the assumed compression in sparticle spectrum. If we accept $\delta a_\mu$ consistency at $3\sigma$, much larger regions of agreement exists for most values of $\tan \beta$. The estimate presented above provides preliminary information about the viability of DMSSM but it should only be regarded as indicative of the true situation. We have used simplified and approximate semi-analytic formulas and assumed the physical and soft masses of sparticles to be the same $\sim M_D$. This approximation is no longer valid in the particular case of large $A$-terms which are necessary here to get the large enough Higgs mass. Further such large $A$-terms can reduce the degeneracy between the sparticles and can even drive some of the sparticle into tachyonic mode in the extreme case. Our estimates of flavor observables are also simplified as they include only the leading order contributions. In order to account for these uncertainties, we provide more accurate numerical analysis of the above scenario in the following section. We will see that some of the results obtained in this section get significantly modified in the next section when full numerical treatment and deviations from exact degeneracy will be considered.

\section{Numerical Analysis} 
\label{numerical}

We now present results from a detailed numerical calculation of the various observables discussed in the previous section. Instead of taking the physical masses of sparticles to be degenerate, we work
with approximate degenerate soft masses and compute the physical mass spectra from it. In order to
account for the various uncertainty and small departure from the exact degeneracy, we allow for each
soft mass and the $\mu$-parameter a random variation within $\pm 10\%$ around the degenerate scale $M_D$, namely
\begin{align}\label{scan-ranges}
m_{\tilde{f}_i} &\in M_D~(1 + \delta_{m_{\tilde{f}}}),\nonumber \\ 
M_{1} = M_{2} = M_{3} &\in M_D~(1 + \delta_M), \\ 
\mu &\in M_D~(1 + \delta_\mu). \nonumber  
\end{align}
The individual delta’s could be different. This essentially makes it a very constrained model with only 5 parameters, {\it i.e.}, $\delta_M$, $\delta_{m_{\tilde{f}}}$, $\delta_\mu$, $M_D$ and $m_A$.  The $\delta_{m_{\tilde{f}}}$, $\delta_M$, and $\delta_\mu$ are independently varied
in the range $[-0.1,\, 0.1]$. The variation is taken to be such that the mass
difference between the gluino and the lightest neutralino 
remains less than 200 GeV when the common degenerate scale approaches to 1 TeV. This mass difference corresponds to the required compactness to escape from the current LHC limits \cite{Bhattacherjee:2012mz,Bhattacherjee:2013wna,Dutta:2015exw}. Due to the choice of the parameters just mentioned the physical masses of the sparticles all lie within a narrow range.

\begin{figure}[t]
\centering
\subfigure{\includegraphics[width=0.45\textwidth]{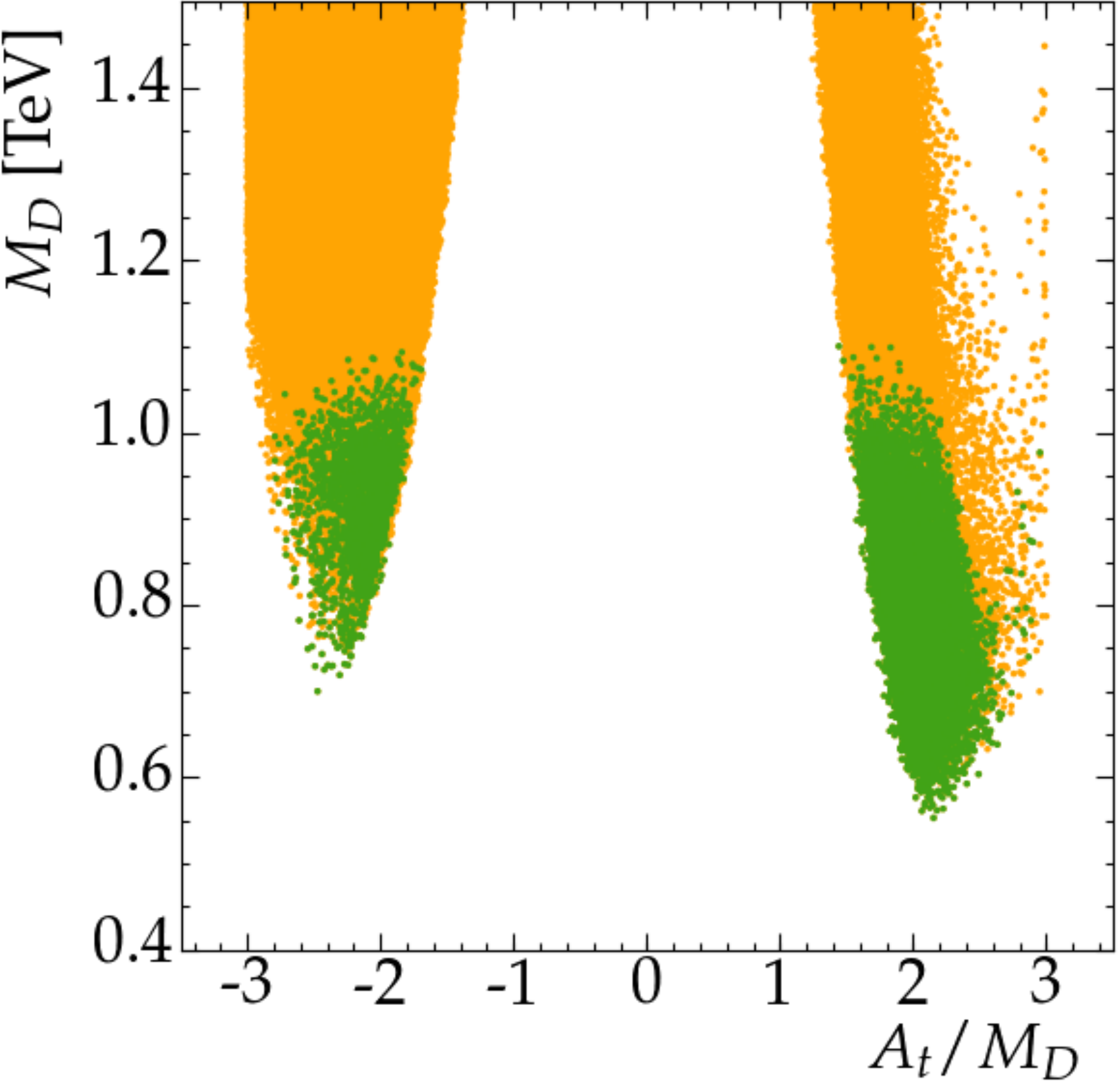}} 
\caption{The regions allowed in $A_t-M_D$ plane for $\pm 10\%$ deviation in the soft masses around the degenerate scale and after imposing the constraints in Eq.~\eqref{indirect-limits}. The green
(orange) points are consistent with the experimental value of $(g-2)_\mu$ at $2\sigma$ ($3\sigma$).} 
\label{fig3}
\end{figure}

We use publicly available package \texttt{SuSeFLAV} \cite{Chowdhury:2011zr} to compute the sparticle 
spectrum and the SUSY contribution to $(g-2)_{\mu}$ at low scale. For the calculation of $B$-physics
observables and dark matter relic density and direct and indirect detection cross-section we use
\texttt{micrOMEGAs 3.2} \cite{Belanger:2013oya}. We calculate all the observables by varying $M_D$ randomly
in the range $[0,\, 1.5]$ TeV, $m_A \in [0.1,\, 10]\, M_D$, $\tan\beta \in [5,\, 60]$, $A_t \in [-3,\, 3]\, M_D$, $ A_b = A_{\tau} = 0$ and the other parameters as specified in Eq.~\eqref{scan-ranges}. We require that the resultant spectrum satisfy the following constraints:
\begin{align}\label{indirect-limits}
m_{h} &\in [122.4,\ 127.8]\ {\rm GeV}, \nonumber \\
 \frac{{\rm BR}(B \rightarrow X_s \gamma)_{\rm MSSM}}{{\rm BR}(B \rightarrow X_s \gamma)_{\rm
 SM}} &\in	 [0.86,\ 1.22]\ (2\sigma)\ \textrm{\cite{Amhis:2014hma}}, \nonumber \\
 {\rm BR}(B_s \rightarrow \mu^+ \mu^-) &\in [1.6,\ 4.2]\times 10^{-9}~(2\sigma)\
 \textrm{\cite{CMS:2014xfa}}, \\
 \frac{{\rm BR}(B^{+} \rightarrow \tau^{+} \nu_\tau)_{\rm MSSM}}{{\rm BR}(B^{+} \rightarrow \tau^{+} \nu_\tau)_{\rm SM}} &\in [0.78,\ 1.78]~(2\sigma)\ \textrm{\cite{Amhis:2014hma}} , \nonumber
\end{align}
as discusses previously. We do not include dark matter constraints here. These will be analyzed separately in the next section. The results for $\pm 10\%$ variation around the degenerate scale $M_D$ are shown in Fig.~\ref{fig3}. The allowed range of $A_t/M_D$ is more limited than in Fig.~\ref{fig2}. This is because Fig.~\ref{fig2} has been made with semi-analytic formulae in section \ref{analytical}, while Fig.~\ref{fig3} uses numerical codes for various computation. The main differences arise from the computation of $m_{h}$. Our numerical analysis uses complete one loop \cite{Pierce:1996zz} and dominant two loop Higgs mass correction which are of $\mathcal{O}\left[ \alpha_s \left( \alpha_t +  \alpha_t \right) + \left(\alpha_t +  \alpha_t \right)^2 + \alpha_{\tau} \alpha_b + \alpha_{\tau}^2 \right]$ \cite{Degrassi:2001yf,Brignole:2001jy,Dedes:2002dy,Dedes:2003km}.

\begin{figure}[t]
\centering
\subfigure{\includegraphics[width=0.46\textwidth]{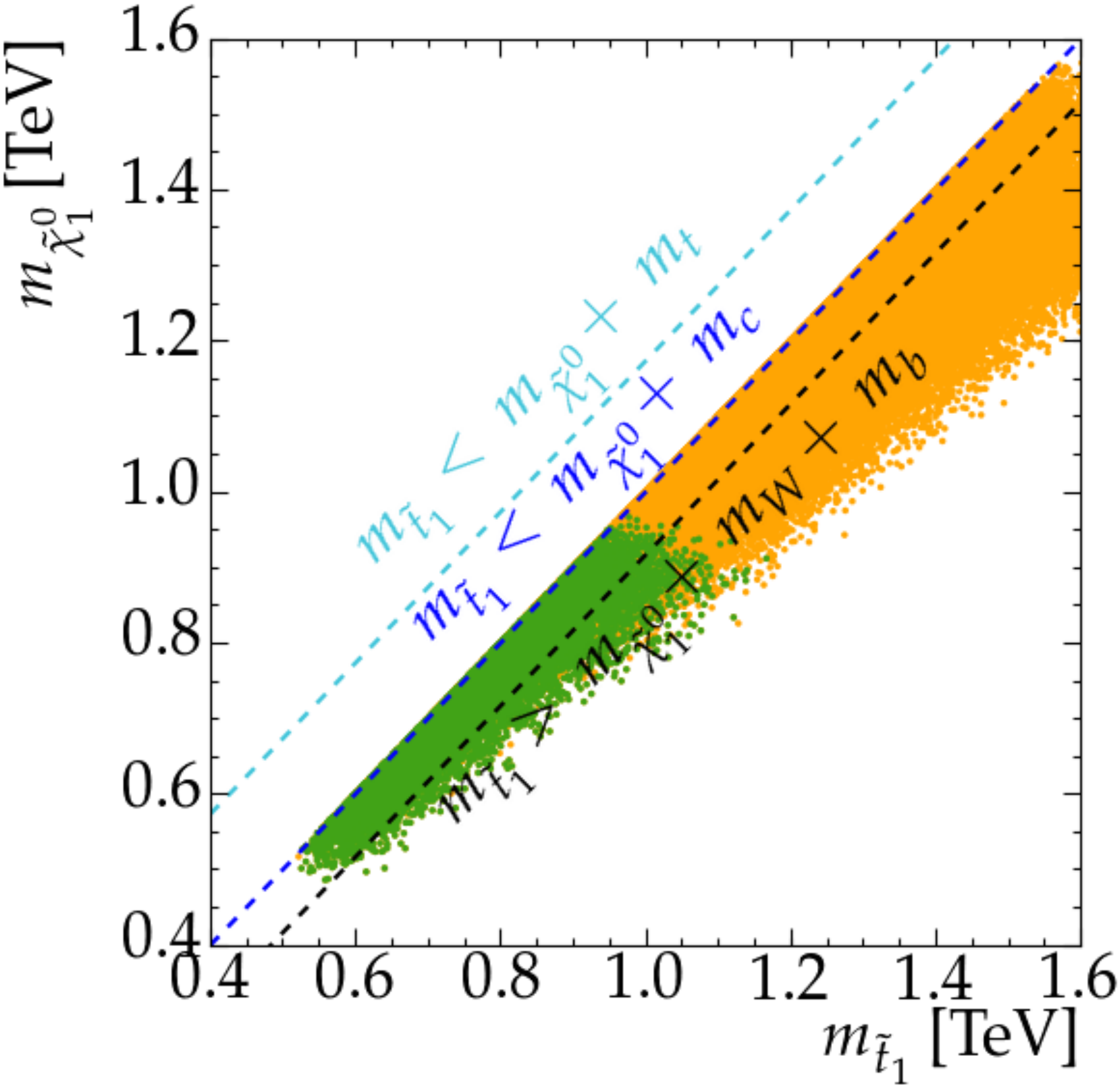}}\quad
\subfigure{\includegraphics[width=0.445\textwidth]{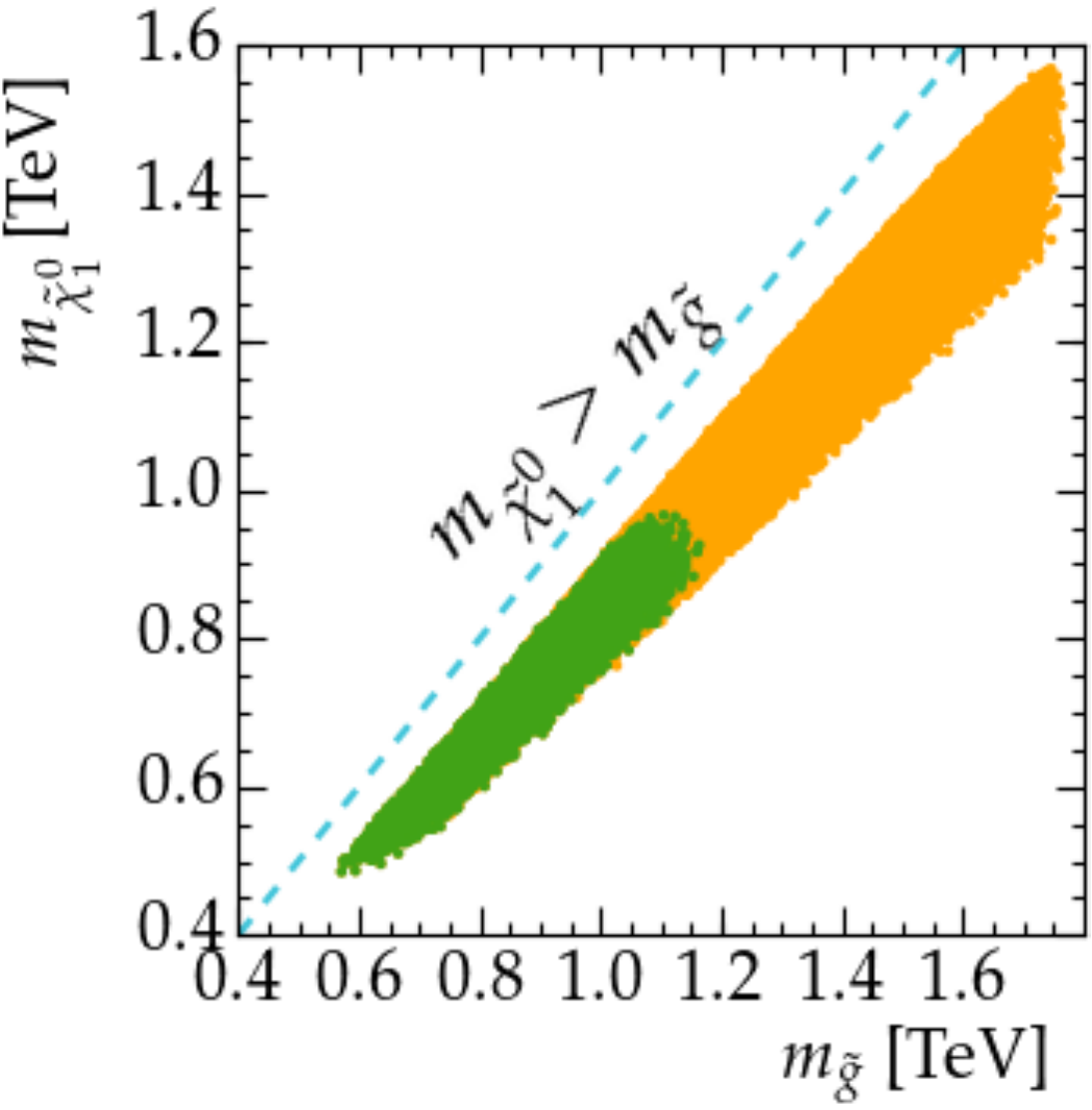}}
\caption{The correlation between the lightest stop (left panel) and the gluino (right panel) with respect to lightest neutralino mass for $\pm 10\%$ deviation in the soft masses around the
degenerate scale and after imposing the constraints in Eq.~\eqref{indirect-limits}. The green
(orange) points are consistent with the experimental value of $(g-2)_\mu$ at $2\sigma$ ($3\sigma$).} 
\label{fig4}
\end{figure}
We note that the $\amu$ constraint, by itself, can be satisfied at $2\sigma$ level by $M_D$ values as low as around $300$ GeV. However, the CP-even Higgs boson is then too light. 
A larger deviation of $m_h$ allows for even lighter spectrum of charginos, neutralinos and sleptons and leading to even smaller $\amu$. Requiring its mass to be in the range given in Eq.~\eqref{indirect-limits}, we find $M_D \gtrsim 500$ GeV as seen from Fig.~\ref{fig3}. Moreover, for this range of $M_D$, the Higgs mass constraint can only be satisfied with large $A_t$. However, such a large $A_t$ enhances the Higgsino-stop loop contribution in the $B\to X_s \gamma$ decay (see Eqs.~(\ref{c7c8},\ref{Rbsg})) significantly. This in turn pushes the degenerate scale to the higher values seen in Fig.~\ref{fig3}. The $B_s \to \mu^+ \mu^-$ constraint remains sub-dominant  in whole of the parameter space. After considering all the constraints in Eq.~\eqref{indirect-limits}, the lower bound on the degenerate scale is $M_D \simeq 600$ GeV for $\pm 10\%$ deviation from the exact degeneracy. 

In the left side panel of Fig.~\ref{fig4}, we show the correlation between the masses of lightest stop and lightest neutralino for the points in Fig.~\ref{fig3}. We see that in the resulting spectrum, the lightest stop could be as light as $550$ GeV with lightest neutralino to be around $500$ GeV. As noted earlier, the current limits on $m_{\tilde{t}_1}$ do not apply if $m_{\tilde{t}_1} > 400$ GeV and $m_{\tilde{t}_1} - m_{\tilde{\chi}^0_1} < 200$ GeV. It is also challenging for the next runs of LHC to probe this entire region because of the possible close degeneracy in the stop-neutralino masses. In order to account for the observed value of $(g-2)$ of muon at $2\sigma$, one gets an upper bound on the lightest stop mass and it is required to be $\lesssim 1$ TeV. For $m_{\tilde{t}_1} \simeq 1$ TeV the stop pair production cross-section is $\sim 10$ fb at 14 TeV LHC.

In the right-hand panel of Fig.~\ref{fig4} we plot the same points as in the left panel but in the $m_{\tilde{g}}-m_{\tilde{\chi}^0_1}$ plane. We see that $m_{\tilde{g}} \lesssim 1.2$ TeV if muonic $(g-2)$ is to be within $2\sigma$ of its measured value. More interesting is the non-vanishing gap between the LSP mass and $m_{\tilde{g}}$. We have checked that this occurs because radiative corrections typically increase $m_{\tilde{g}}$ by a factor $\sim \frac{15\alpha_3}{4 \pi} \sim 10\%$, while mixing effects tend to reduce the mass of the LSP as well as the lighter top squark. The qualitative difference in the stop-LSP and gluino-LSP mass gaps (which obviously impact LHC searches) plays an important role in the determination of the neutralino thermal relic density as discussed in section \ref{dm}.

Spurred by the fact that some limits on the gluino mass in Ref.~\cite{ATLAS-CONF-2016-037,ATLAS-CONF-2016-052,ATLAS-CONF-2016-054,ATLAS-CONF-2016-078,CMS-PAS-SUS-16-014,CMS-PAS-SUS-16-016,CMS-PAS-SUS-16-022,CMS-PAS-SUS-16-030,CMS-PAS-SUS-16-020,CMS-PAS-SUS-16-019} appear to be valid our to $m_{\tilde{g}}=900$ GeV and $m_{\tilde{g}}-m_{\tilde{\chi}^0_1}$ is as small as $100$ GeV which nearly seems to exclude some points in the right frame of Fig.~\ref{fig4}, we have examined these exclusions more carefully. For the most part, these come from analyses in simplified models with heavy squarks where it is assumed that $\tilde{g} \rightarrow q\bar{q}\tilde{\chi}_1^0$, $b\bar{b}\tilde{\chi}_1^0$, or $q\bar{q}W^{(*)}\tilde{\chi}_1^0$ with a branching fraction of $100\%$, and requiring up to six jets in the event \cite{Aaboud:2016zdn}.  Clearly the analysis requiring tagged $b$-jets is applicable only to a fraction of points in the figure. Moreover, in the DMSSM with the squark heavier than the gluino, the gluino decays are split into several chargino and neutralino modes, but more importantly, the squark is typically close in mass to $m_{\tilde{g}}$ so that the daughter quark is relatively soft because the squark is dynamically preferred to be close to on-shell, suppressing events with high jet multiplicities. We expect, therefore, that the gluino mass bound is then substantially reduced from the value of $\sim 700$ GeV in Fig.~7 of Ref.~\cite{Aaboud:2016zdn} for a compressed spectrum. If, on the other hand, the squark is lighter than the gluino, gluino can decay to squarks, and bounds for squarks degenerate with the LSP are significantly weakened for masses larger than 500-600 GeV, even assuming all squarks decay directly to the LSP. The non-observation of an excess of monojet events at the LHC can be translated into an independent lower limit, $m_{\tilde{q}} > 550-600$ GeV ~\cite{Aaboud:2016tnv}, on the squark mass, assuming again that squarks directly decay to the LSP. This limit will again be weakened in the DMSSM. The point of this discussion is not that \textit{all} the points in the DMSSM scan in Fig.~\ref{fig4} survive the LHC bounds, but simply that a large fraction of these survive and furthermore that many of these are consistent with the observed value of $(g-2)_{\mu}$. Specialized strategies would be needed to thoroughly probe the DMSSM parameter space.

\begin{figure}[t]
\centering
\includegraphics[width=0.9\textwidth]{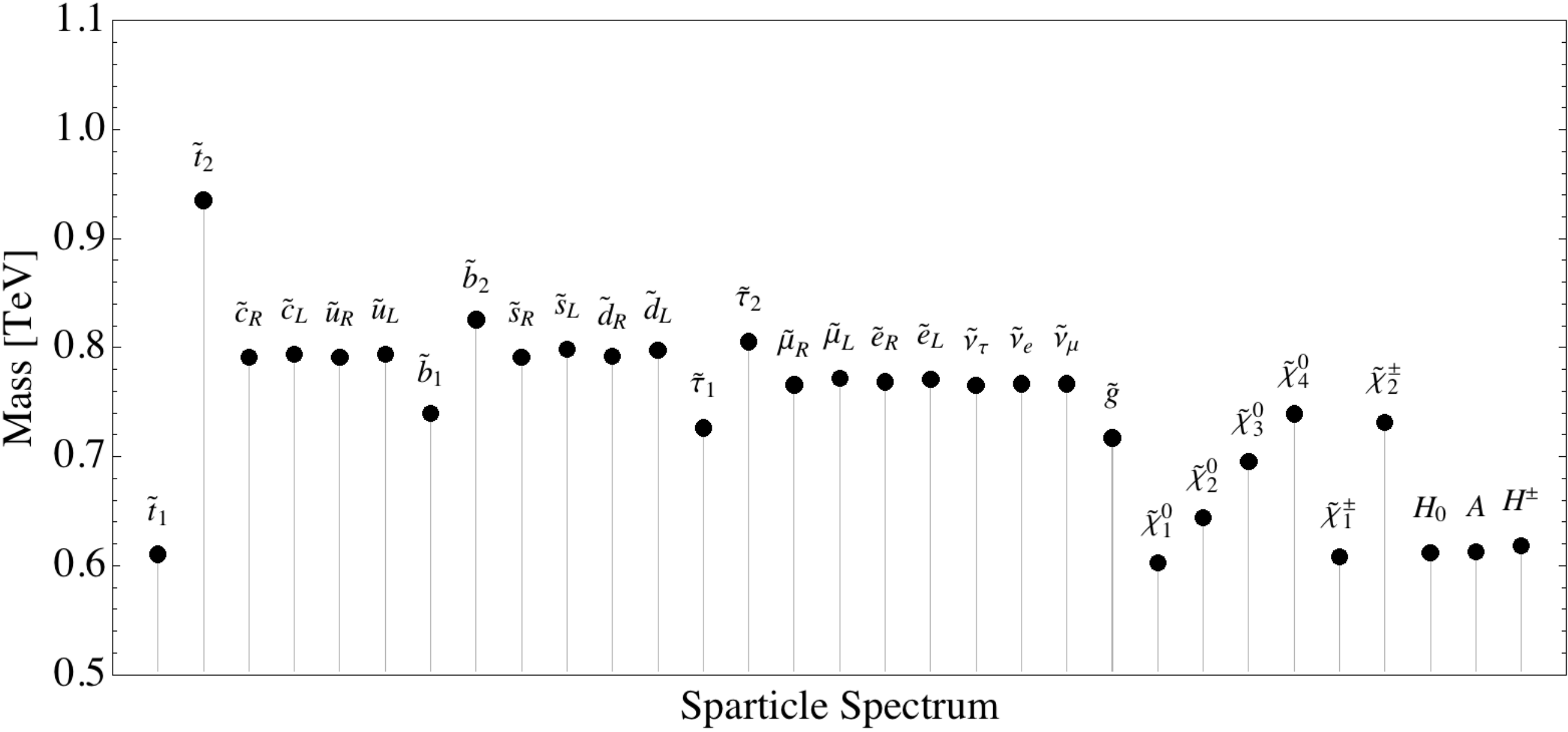}
\caption{Typical spectrum in the case of $\mgmt \simeq 700$ GeV, which satisfies the bounds in 
Eq.~\eqref{indirect-limits}. This point has $a_\mu = 1.5 \times 10^{-9}$ .}
\label{sp-1}
\end{figure}

\begin{figure}[t]
\centering
\includegraphics[width=0.9\textwidth]{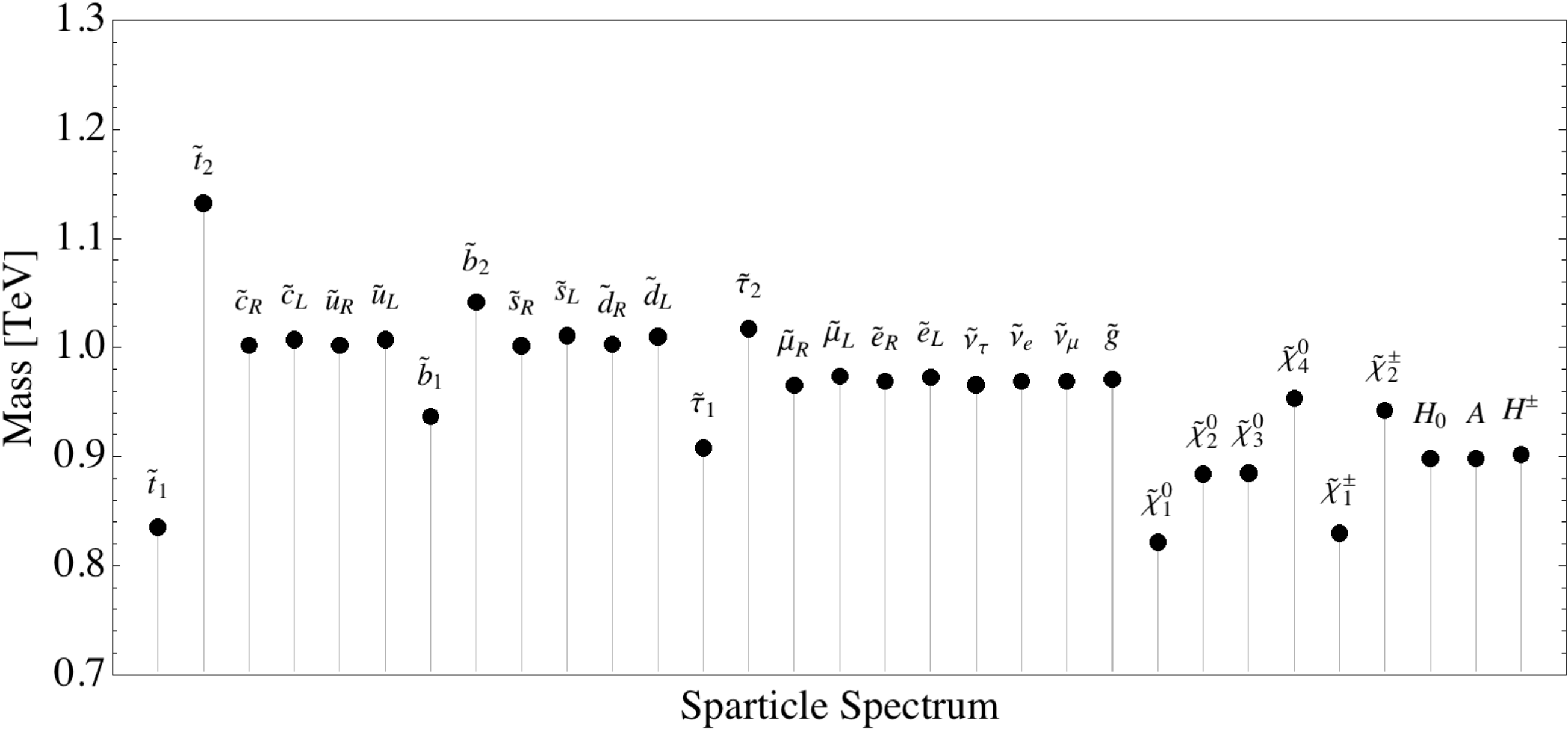}
\caption{Typical spectrum in the case of $\mgmt \simeq 975$ GeV, which satisfies the bounds in
Eq.~\eqref{indirect-limits}. This point has $a_\mu = 1.18 \times 10^{-9}$.}
\label{sp-2}
\end{figure}

In Figs.~\ref{sp-1} and \ref{sp-2}, we show the spectrum for two example benchmark points  which fall in the green regions of Fig.~\ref{fig3}. Notice that for these benchmark points the sleptons (and squarks) are heavier than electroweak-inos. This is important because (unless squarks are also light), the -inos may decay leptonically 100\% of the time and be in conflict with LHC data \cite{ATLAS-CONF-2016-096}. For charginos and neutralinos whose branching fractions mirror those of $W$ and $Z$ bosons, the LHC lower limits are $\lesssim 400$ GeV even for a massless LSP.  

The results of our numerical analysis are in good agreement with the semi-analytic results obtained in the exact degenerate limit in the previous section. As it can be seen from Fig.~\ref{fig2}, the exact degeneracy requires $M_D$ in the range $800-1000$ GeV and large $A_t$ as well as $\tan\beta$ in order to resolve the muonic $(g-2)$ discrepancy within $2\sigma$ while being consistent with the other direct and indirect constraints. Deviating slightly from the exact degeneracy limit significantly releases the lower bound on $M_D$ and one can have the degenerate scale as low as 550 GeV consistent with all the constraints considered in this paper. The major difference between the two approaches is that we allow a significantly wider range in $m_A \in [0.1,\, 10]\, M_D$ in the numerical analysis and do not consider it to be degenerate with $M_D$ as it is assumed in the semi-analytic study. The large value of $m_A \sim m_{H^+}$ suppresses the charged Higgs and pseudoscalar mediated contributions to $B \to X_s \gamma$ and allowing more room for light $M_D$ as displayed in Fig.~\ref{fig3}. This relatively large $m_A$ also helps in evading $B_s \to \mu^+ \mu^-$ constraints and allows more space in $A_t$-$M_D$ plane compared to that in Fig.~\ref{fig2}. Also note that the disparity with respect to the sign of $A_t$ also becomes feeble and even negative values of $A_t$ are allowed when the degeneracy between $m_A$ and $M_D$ is removed. We find that when $m_A \approx M_D$ is imposed in the numerical analysis, the $M_D$ is pushed to 900 GeV which is then in very good agreement with the results of semi-analytic studies. Clearly, the constraints from $B$-physics can be relaxed significantly when the degeneracy in some masses are relaxed and the results of Fig.~\ref{fig2} get modified.

\section{Dark matter}
\label{dm}
We now turn to constraints on the DMSSM that arise from the measurement of the cold dark matter relic density,
\begin{align}\label{planck}
0.1131 < \Omega_{\rm CDM}\, h^2  < 0.1263\ (3\sigma)\ \textrm{\cite{Planck:2015xua}},  
\end{align}
by the Planck collaboration, assuming that the thermally produced neutralino forms all or part of the observed dark matter in the present universe. Since the dark matter could consist of several components we interpret the Planck measurement given in Eq.~\eqref{planck} as the constraint $\Omega_{\schi^0_1}\, h^2  < 0.1263$.

\begin{figure}[ht]
\centering
\subfigure{\includegraphics[width=0.48\textwidth]{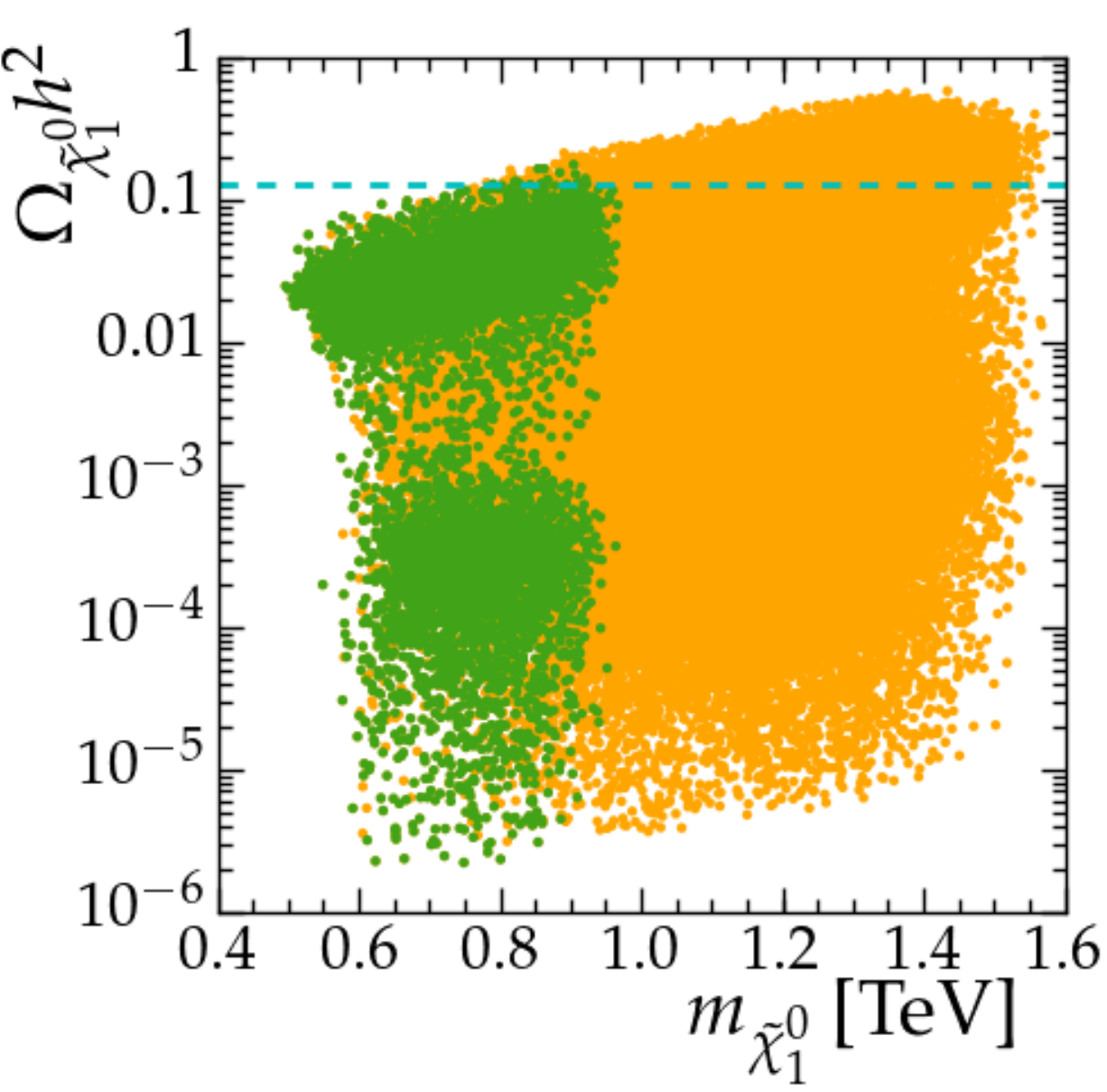}} 
\caption{Results of a scan of the neutralino relic density versus its mass for $\mgmt \in [0,1500]$ GeV with $\pm 10 \%$ deviation in the soft masses from the exact degeneracy for scan points that satisfy all the constraints in Eq.~\eqref{indirect-limits}. The horizontal dashed line shows the upper bound on the dark matter relic density given in Eq.~\eqref{planck}.}
\label{dm-om}
\end{figure}

We begin by evaluating the neutralino thermal relic density for the points from Fig.~\ref{fig3} which survive the various direct and indirect constraints described in the previous section. The results are displayed in Fig.~\ref{dm-om}.
We see that the LSP remains under-abundant over most of the parameter space. This is because in the DMSSM, the LSP is either dominantly higgsino-like, or is an admixture of higgsinos, bino, and even wino. In these cases it is well-known that the LSPs rapidly annihilate to $W^{+}W^{-}$ pairs in the early universe (co-annihilations with the charginos may also be important \cite{Mizuta:1992qp}), resulting in a thermal neutralino relic abundance well below the measured value $\Omega_{\rm CDM}\, h^2 \simeq 0.12$.

Before moving further, we note that $\Omega_{\schi^0_1}\, h^2$ assumes values below $10^{-5}$ for points in our scan with $m_{\schi^0_1} \sim 1$ TeV. This is three orders of magnitudes below the naive expectation of $\Omega_{\widetilde{W}}\, h^2 \sim 0.1 (\frac{m_{\widetilde{W}}}{3\textrm{ TeV}})^2$ , for the relic density of thermally produced winos,\footnote{If the neutralinos has significant higgsino or bino components, the expected neutralino density would be even larger.} obtained by scaling the annihlation cross-section as $\frac{1}{m_{\widetilde{W}}^2}$ and remembering that 3 TeV winos saturate the observed CDM relic density in Eq.~\eqref{planck}. We attribute this to the importance of neutralino co-annihilation with strongly interacting superpartners, most notably stops \cite{Boehm:1999bj,Ellis:2001nx,Raza:2014upa} which, as we see in Fig.~\ref{fig4} (left panel), can be nearly degenerate with the LSP in our model, but not in most usually considered SUSY models (where colored superpartners are much heavier than the LSP). We have checked that co-annhilation with the gluino \cite{Profumo:2004wk,Chen:2010kq,Harigaya:2014dwa,Raza:2014upa,Ellis:2015vaa,Ellis:2015vna} does not play a big role because radiative corrections typically increase the $\tilde{g}-\tilde{\chi}_1^0$ mass gap (see Fig.~\ref{fig4}, right panel), leading to a Boltzmann suppression of the number density of the gluinos in the early universe.

\begin{figure}[t]
\centering
\subfigure{\includegraphics[width=0.48\textwidth]{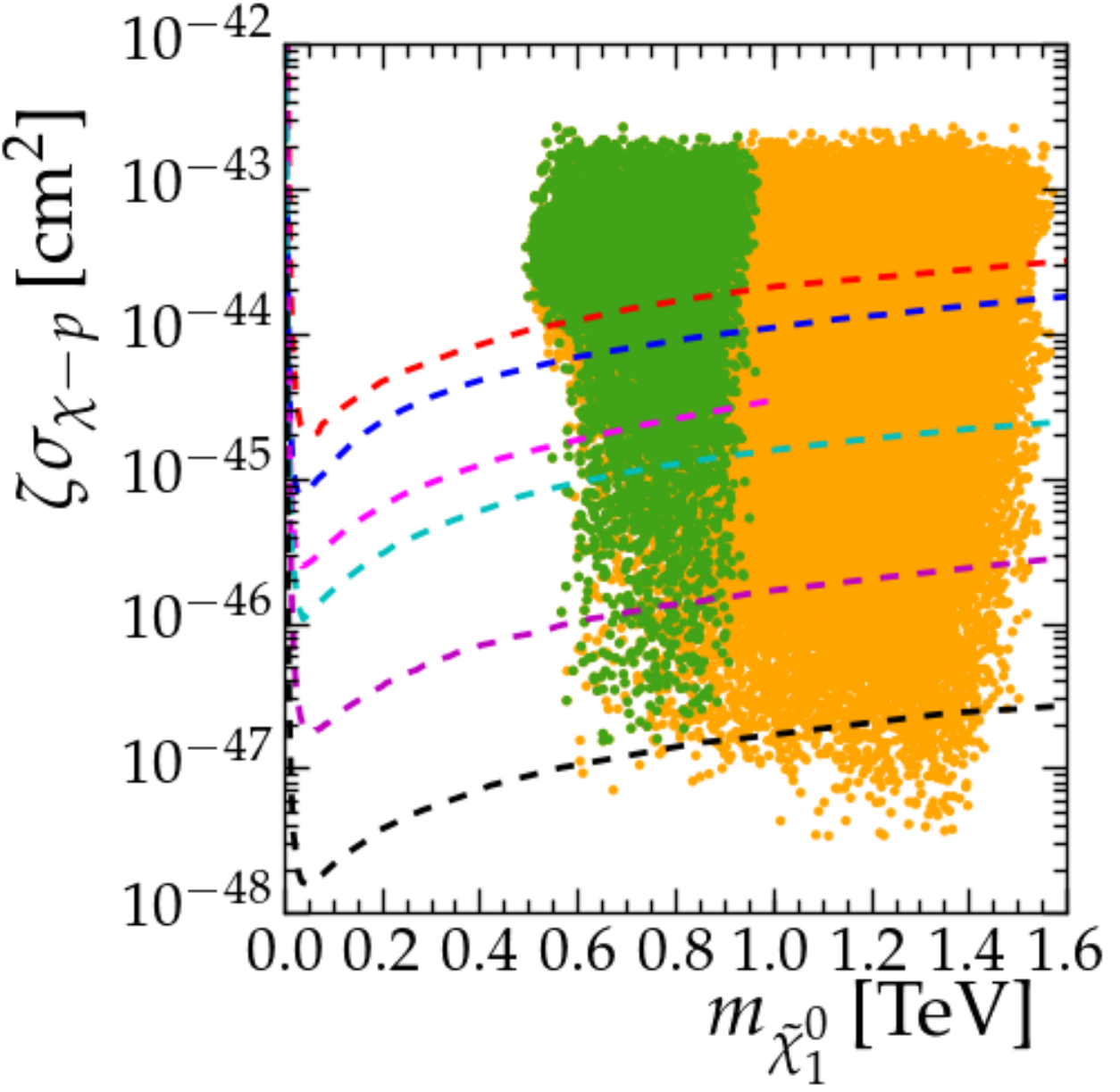}}\quad
\subfigure{\includegraphics[width=0.48\textwidth]{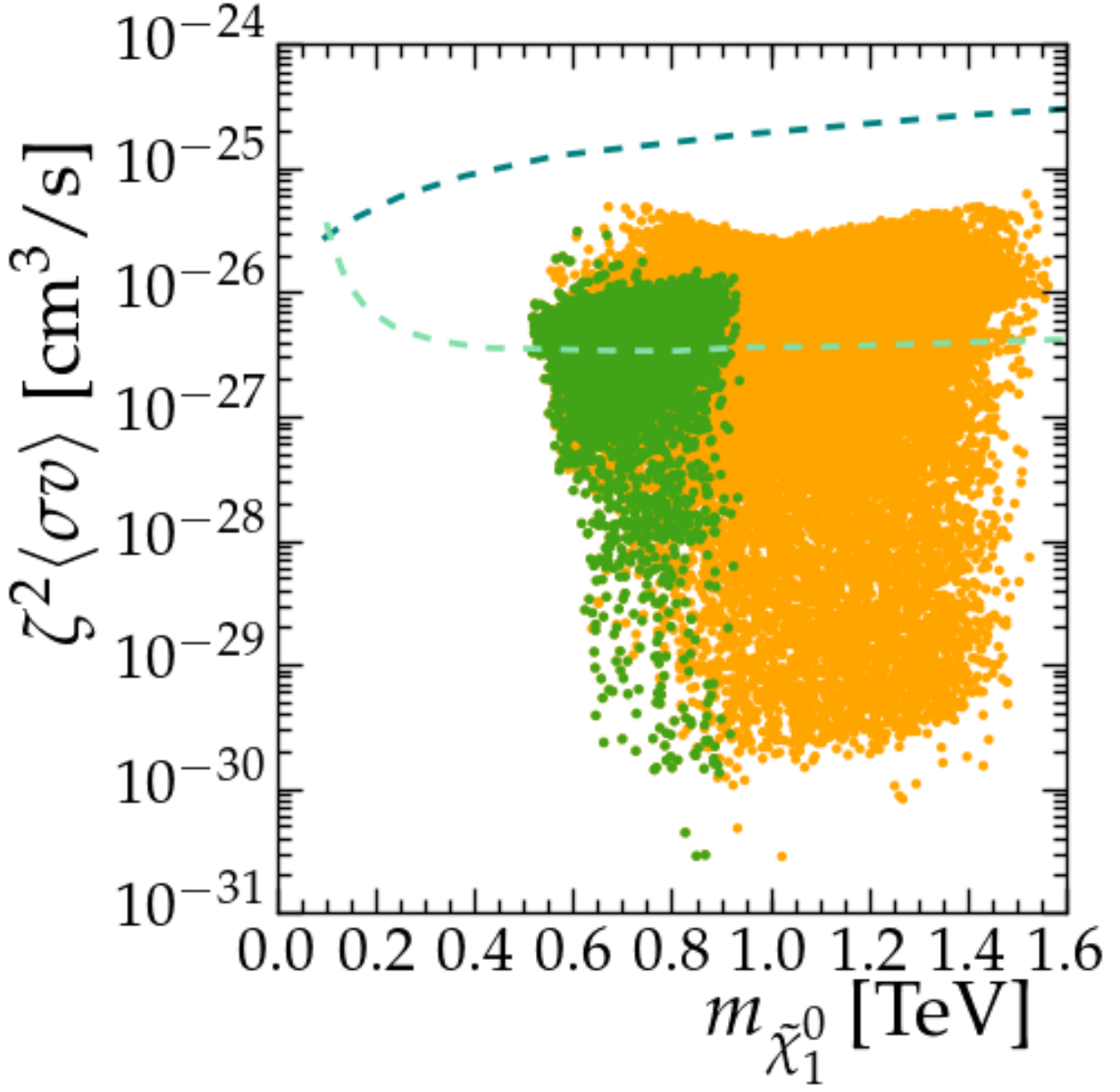}}
\caption{Left: For the same points in Fig.~\ref{dm-om}, the scaled neutralino-nucleon spin-independent cross-section with respect to the lightest neutralino mass. In both panels, the green (orange) points are in agreement with the experimental value of the muon $g-2$ at $2\sigma$ ($3\sigma$). The red-dashed and blue-dashed line correspond to the bounds by the XENON100 \cite{Aprile:2012nq} collaboration and from the data taken in 2013 by the LUX \cite{Akerib:2013tjd} collaboration. The pink-dashed (cyan-dashed) line represents the latest limit from PandaX-II \cite{Tan:2016zwf} (LUX \cite{Akerib:2016vxi}) collaboration. The magenta (black) dashed line shows the projected reach of the XENON1T (XENONnT) \cite{Aprile:2012zx,Aprile:2015uzo} collaboration. Right: Variation of the scaled neutralino annihilation cross-section with respect to the neutralino mass for the points in Fig.~\ref{dm-om}. The dark cyan dashed line represents the upper bound on DM annihilation to $W^{+}W^{-}$ pairs from Fermi-LAT and MAGIC collaboration \cite{Ahnen:2016qkx} from the search of gamma-ray signals in dwarf satellite galaxies while the light green dashed line depicts the projected reach of the CTA \cite{Wood:2013taa} from gamma-ray searches.}
\label{dm-dd-id}
\end{figure}
Next we turn to prospects for dark matter detection and constraints from on-going experiments. Toward this end, we evaluate the spin-independent neutralino-nucleon scattering cross-section and compare it with data from XENON100 \cite{Aprile:2012nq}, PandaX-II \cite{Tan:2016zwf}, and LUX \cite{Akerib:2013tjd,Akerib:2016vxi}, as well as make projections for reach of the XENON1T \cite{Aprile:2012zx} experiment. Spin-independent neutralino-nucleon scattering arises from $t$-channel Higgs boson exchange as well as from squark exchanges in the $s$- and $u$-channels. Since the neutralino relic density in Fig.~\ref{dm-om} is for the most part too low,  we assume that the neutralino forms only one component of the dark matter. In this case, to get the correct estimate for the neutralino-nucleon event-rate in direct detection experiments, one should scale the neutralino-nucleon cross-section calculated assuming neutralino as the single component CDM by the fraction, $ \rho_{\schi^0_1}/\rho_0$, where $\rho_0$ denotes the total local dark matter density and $\rho_{\schi^0_1}$ is the dark matter density contributed by the neutralino. In the left panel of Fig.~\ref{dm-dd-id}, we show the neutralino-nucleon spin-independent scaled cross-section versus the  neutralino mass for all the points in Fig.~\ref{fig3}, where $\zeta = {\rm min}(1, \Omega_{\schi^0_1}\, h^2 / \left. \Omega_{\rm CDM}\,h^2\right|_{\rm min})$. By definition for a single component dark matter or for correct relic abundance $\zeta$ is unity. From Fig.~\ref{dm-om}, it can be seen that current limits from the XENON100, LUX, and PandaX-II experiment, rules out a considerable part of the green region consistent with the $(g-2)_{\mu}$ at $2\sigma$ level, whereas the orange region, satisfying the $(g-2)_{\mu}$ at $3\sigma$, is less constrained. It is important to note that the residual orange and green regions will be probed by the future direct-detection experiments like XENON1T and its upgrades \cite{Aprile:2015uzo}. 

In the right panel of Fig.~\ref{dm-dd-id}, we show the properly scaled thermally averaged neutralino cross-section to $W^+W^-$ pairs times the relative velocity of the neutralinos for the same points as in left panel of Fig.~\ref{dm-dd-id}. The dark cyan dashed line corresponds to the upper bound on the DM annihilation cross-section to $W^+W^-$ pairs from the combined analysis of gamma-ray data from the Fermi-LAT and MAGIC collaborations \cite{Ahnen:2016qkx}  searching for gamma-ray signals from dark matter annihilation in dwarf satellite galaxies. We see that this analysis does not lead to any additional constraints largely because the expected event rate from neutralino annihilation scales as $\zeta^2$. A future ground-based gamma-ray observatory like Cherenkov Telescope Array (CTA) \cite{Wood:2013taa} will be able to probe some parts of the parameter space. We have shown its sensitivity by the green line assuming 500 hours of exposure.

\section{Summary}
\label{summary}
If the masses of superpartners of the SM particles are nearly degenerate, they can easily escape detection at the LHC even if they are relatively light. Indirect limits coming from various flavor violating effects are also evaded. Light superparticles can alleviate the discrepancy between the SM prediction and the experimental value of muon magnetic moment, but of course, cause tension with the observed value of the Higgs boson mass. Motivated by this, we present a phenomenological study of the DMSSM scenario in which all the soft masses as well as the higgsino mass $\mu$ are considered approximately degenerate. 

We analyzed the viability of such a spectrum in view of the various direct and indirect constraints from the Higgs boson mass, $B$-physics, muon $g-2$, dark matter, and of course, the non-observation of signals at the LHC. In our phenomenological analysis, we set the magnitudes of all soft mass parameters and also the $\mu$-parameter to a common mass scale $\sim M_D$ and allow a small and independent departures, within $\pm 10\%$, from exact degeneracy: see Eq.~\eqref{scan-ranges}. The resultant sparticle spectrum is very compact and consistent with the limits obtained by the ATLAS and CMS collaborations. Our analysis shows that the DMSSM can account for the measured value of muon $g-2$ at or less than 2$\sigma$, if the degenerate scale is $M_D$ is in the range of $300 - 500$ GeV ($500 - 1100$ GeV) for small (large) $\tan\beta$. On the other hand, the observed Higgs mass requires large negative $A_t$ for low $M_D$ values leading to sizable flavor violations in $B$-decays, particularly in $B \to X_s \gamma$ channel. The required suppression in flavor violations then drives the degenerate scale towards the higher values. The constraints from $B \to X_s \gamma$ and $B_s \to \mu^+ \mu^-$ can be relaxed if the pseudoscalar and charged Higgs masses are taken higher than $M_D$.

We find the range of $M_D \in [600, 1000]$ GeV that can explain the experimental measurement of muon $(g-2)$ at 2$\sigma$ respecting all the relevant direct and indirect constraints considered in this paper. The solution prefers large $\tan\beta$ as well as a large trilinear term, $A_t$. The physical mass spectrum is compact with gluino and light stop masses smaller than $\sim 1.1$ TeV, and corresponding mass differences with the LSP $\lesssim 250$ GeV, even at the highest sparticle masses\footnote{Allowing a $3\sigma$ difference for muonic $(g-2)$ will allow significantly heavier superpartners.}.  Such stops and gluinos would be kinematically accessible at the LHC, but specialized strategies would be needed for their detection above SM backgrounds. We have also examined dark matter in the DMSSM. We show that the lightest neutralino could form part, but not all, of the dark matter, primarily because it is typically a mixed bino-higgsino-wino state that would annihilate very efficiently in the early universe. Much of the parameter space where muonic $(g-2)$ is within $2\sigma$ of its measured value is excluded by the direct search limits from the XENON100, PandaX-II and LUX experiments while most of the remaining space will be probed by XENON1T, and essentially completely by the XENONnT experiment \cite{Aprile:2015uzo}. Indirect searches for DM seem to be much less limiting for our scenario.

In summary, we have shown that the DMSSM allows for a SUSY explanation of the muon anomalous magnetic moment while at the same time satisfying all current experimental bounds. Squarks and gluinos should be abundantly produced at the LHC and it remains a challenge to develop specialized strategies to isolate their signals from SM backgrounds. Direct detection DM searches offer an alternative way to probe this otherwise difficult scenario.

\begin{acknowledgments}
We acknowledge useful discussions with Biplob Bhattacherjee, Miko\l{}aj Misiak, and Shahram Rahatlou. The work of DC has been supported by the European Research Council under the European Union's Seventh Framework Programme (FP/2007-2013)/ERC Grant Agreement n.~279972. KMP thanks the Department of Science and Technology, Government of India for research grant support under INSPIRE Faculty Award (DST/INSPIRE/04/2015/000508).  XT was supported, in part, by the U.S. Department of Energy Grant No.~DE-SC-0010504. SKV acknowledges support from IUSTFF Grant: JC-Physics Beyond Standard Model/23-2010 and thanks Department of Physics and Astronomy, University of Hawaii for hospitality where this work was started. 
\end{acknowledgments}

\bibliographystyle{apsrev4-1} 
\bibliography{csusy-gm2}
 
\end{document}